\pdfoutput=1

\documentclass[11pt,draftcls,onecolumn]{IEEEtran}
\usepackage[T1]{fontenc}
\usepackage{amsmath} 
\usepackage{amssymb} 
\usepackage[final]{graphicx} 
\usepackage{color}
\usepackage{hyperref}

\newcommand{\mathscr}{\mathcal}

\DeclareMathOperator{\Aut}{Aut}
\newcommand{\bigast}{\mbox{\Large $*$}} 
\newcommand{\bsy}[1]{\boldsymbol{#1}} 
\newcommand{\dd}{\ensuremath{\check{d}}}  
\newcommand{\diag}{\mbox{\rm diag}}

\newcommand{\dotcup}{\ensuremath{\mathaccent\cdot\cup}} 
\newcommand{\gam}[1]{\mbox{$\gamma_{\!\ssst#1}$}} 
\newcommand{\order}{\mbox{\rm order}}
\DeclareMathOperator{\sgn}{sgn}
\newcommand{\ssst}{\scriptscriptstyle}

\newcommand{\thet}[1]{\mbox{$\theta_{\!\ssst#1}$}} 

\newtheorem{cor}{Corollary} 
\newtheorem{defn}{Definition} 
\newtheorem{exmp}{Example} 
\newtheorem{lem}{Lemma} 
\newtheorem{prop}{Proposition} 
\newtheorem{thm}{Theorem}
\newcommand{\case}[1]{\paragraph*{Case}#1\  } 
\newcommand{\rem}{\paragraph*{Remarks}} 
\newcommand{\pf}{\IEEEproof} 
\newcommand{\qed}{\hfill\IEEEQED} 

\begin{document}
\date{September 29, 2013}
\title{Group-Theoretic Structure of Linear Phase Multirate Filter Banks}
\author{Christopher M.\ Brislawn
\\Los Alamos National Laboratory, MS B265, Los Alamos, NM 87545--1663 USA\\
(505) 665--1165 (office);\quad  (505) 665--5220 (FAX);\quad e-mail: {\tt brislawn@lanl.gov}
\thanks{The author is with Los Alamos National Laboratory, Los Alamos, NM 87545--1663 USA.  Los Alamos National Laboratory is operated by Los Alamos National Security LLC for the U.\ S.\ Department of Energy under contract DE-AC52-06NA25396. This work was partially supported by the Los Alamos Laboratory-Directed Research \& Development Program, Kristi~D.\ Brislawn, and Reilly~R.\ Brislawn.}
}

\IEEEaftertitletext{\vspace{-1\baselineskip}%
\centerline{\em Dedicated to Professor Arlan B. Ramsay on his retirement.}
}

\maketitle
\vspace{1\baselineskip}
\begin{center}{\bf  FINAL PREPRESS CORRECTIONS (rev. 4), 9/29/13, POSTED TO arXiv.org}\end{center}
\vspace{1\baselineskip}

\begin{abstract}
Unique lifting factorization results for  group lifting structures are used to characterize the group-theoretic structure of  two-channel   linear phase FIR perfect reconstruction filter bank groups.  For $\mathscr{D}$-invariant, order-increasing group lifting structures, it is shown that the associated lifting cascade group $\mathscr{C}$ is isomorphic to the free product of the upper and lower triangular lifting matrix groups.  Under the same hypotheses, the associated scaled lifting group $\mathscr{S}$ is  the semidirect product of  $\mathscr{C}$ by the diagonal gain scaling matrix group $\mathscr{D}$.  These results apply to the group lifting structures for the two principal classes of  linear phase perfect reconstruction filter banks, the whole- and half-sample symmetric classes.  Since the unimodular whole-sample symmetric class forms a group,  $\mathscr{W}$, that is in fact equal to its own scaled lifting group,  $\mathscr{W=S_W}$, the results of this paper characterize the group-theoretic structure of   $\mathscr{W}$ up to isomorphism.  Although the half-sample symmetric class  $\mathfrak{H}$ does not form a group, it can be partitioned  into cosets of  its  lifting cascade group, $\mathscr{C}_\mathfrak{H}$, or, alternatively, into cosets of  its scaled lifting  group, $\mathscr{S}_\mathfrak{H}$.  Homomorphic comparisons reveal that scaled lifting groups covered by the results in this paper have a structure analogous to a ``noncommutative vector space.''
\end{abstract}


\newpage\tableofcontents\listoffigures\newpage

\section{Introduction}\label{sec:Intro}
Finite impulse response (FIR) multirate filter banks have become important  tools  in a variety of digital audio and image coding applications.  Perfect reconstruction (PR) filter banks are invertible linear transformations and are typically employed in subband coding schemes that split their sources  into multiple frequency subbands for encoding and transmission.  This enables subband rate allocation strategies that provide significant coding gain over direct quantization and entropy encoding of untransformed  input.  The present paper studies two-channel  filter banks, which  are commonly cascaded to generate more complicated frequency partitions.     As with Fourier transforms,  filter banks have  corresponding analog  transforms, and  filter bank cascades are often called \emph{discrete wavelet transforms} (DWTs) if the filter bank corresponds to an analog wavelet multiresolution analysis~\cite{Daub92,VettKov95,StrNgu96,Mallat99}.  

The reason  subband coding is often proving superior  to  traditional block transform coding  based on  Fourier, cosine, or Karhunen-Loeve transforms is \emph{localization}.  The subbands produced by a  filter bank are  samplings of the signal's information content that are simultaneously localized  in both time (or space) \emph{and} frequency.  This eliminates the need for block-based or windowed transforms   to achieve  joint time-frequency localization.  Such joint localization allows  frequency-dependent quantization and entropy coding to adapt to nonstationary input.  Moreover, unlike traditional closed-loop   prediction schemes such as differential pulse code modulation (DPCM), FIR filter banks are \emph{open-}loop transforms that allow random access into coded bitstreams by decoding a limited subset of bitstream data nonrecursively.

Nonetheless, Fourier analysis still has a 200-year head start on wavelets and filter bank theory.  
The use of Fourier transforms to analyze arbitrary translation-invariant linear operators is highly evolved.  Fourier analysis has  been defined on arbitrary locally compact abelian groups, and the effort to generalize Fourier analysis to noncommutative settings has led to the theory of unitary representations for nonabelian groups.  Applications of Fourier analysis in science and engineering are widespread.  Even   recent developments like  numerical algorithms and hardware based on  the fast Fourier transform (FFT) enjoy big head starts over algorithmic and hardware developments for filter banks.

\begin{figure}[t]
  \begin{center}
    \includegraphics{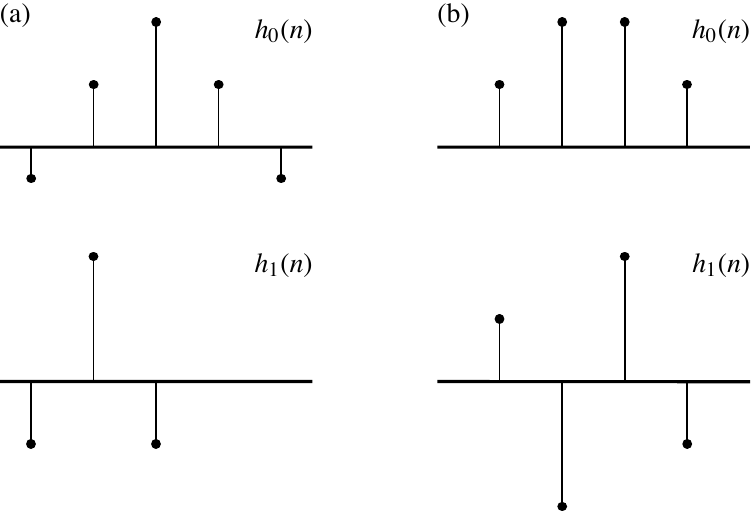}
    \caption{Examples of the two principal classes of linear phase filter banks. (a)~Whole-sample symmetric (WS) filter bank.\quad 
    (b)~Half-sample symmetric (HS) filter bank.}
    \label{WS-HS}
  \end{center}
\end{figure}

The present paper attempts to narrow the  maturity gap between Fourier and multiresolution analysis  a little bit by using established mathematics to characterize the algebraic structure  of multirate filter banks.  Rather than generalizing  to highly abstract settings or    exotic filter banks, we concentrate on developing a deeper understanding of filter banks  that have already proven their value in practical applications, namely, two-channel linear phase FIR PR filter banks.  The \emph{whole-sample symmetric} (WS) and \emph{half-sample symmetric} (HS) classes, whose  highpass impulse responses are, respectively,  symmetric or antisymmetric (see Fig.~\ref{WS-HS}), correspond to multiresolution analyses with compactly supported symmetric or antisymmetric mother wavelets.  We will show that these two principal classes of linear phase filter banks  can be described in  detail using   group theory, something that has not been done previously.  Our primary tool  is the uniqueness theory for  lifting factorizations developed  in~\cite{Bris:10:GLS-I,Bris:10b:GLS-II} and outlined in~\cite{Bris:13:FFT}.

\subsection{Outline of the Paper}\label{sec:Intro:Outline}
Section~\ref{sec:Coding} presents background on  communication  coding standards involving   filter banks so  readers can appreciate the  role filter banks play in contemporary  digital communications  and can assess the evolving state of the art.  
Section~\ref{sec:Review} briefly reviews    notation and terminology from \cite{Bris:10:GLS-I,Bris:10b:GLS-II} regarding group lifting structures.  
Section~\ref{sec:Free} identifies lifting cascade groups that have unique irreducible group lifting factorizations~\cite[Theorem~1] {Bris:10:GLS-I}  with  free products of  lower and upper triangular lifting matrix groups.  
Section~\ref{sec:Semidirect} presents the semidirect product representation of  scaled lifting groups, a result that follows easily, via an independent argument, from the  same hypotheses as  \cite[Theorem~1] {Bris:10:GLS-I}.  This provides the group-theoretic characterization of the WS filter bank group.  The HS class, which does \emph{not} form a group, is completely described in terms of \emph{cosets} of its associated matrix groups.  
Section~\ref{sec:Comparison} shows  how the group-theoretic structure of scaled lifting groups parameterizes linear phase filter banks in terms of a unique factorization framework that structurally enforces  perfect reconstruction and linear phase properties.  A homomorphic correspondence  between the formal algebraic properties of scaled lifting groups and \emph{most} of the axioms for vector spaces exhibits scaled lifting groups as a type of ``noncommutative vector space'' (i.e., a nonabelian group with a group of scaling automorphisms).


\section{Background on Multirate Filter Banks in Digital Coding Standards}\label{sec:Coding}

\subsection{Speech Coding}\label{sec:Coding:Speech}

\subsubsection{G.711}\label{sec:Coding:Speech:G.711}
International standards for narrowband digital speech coding based on the venerable A-law and $\mu$-law logarithmic companding algorithms date back to ITU-T Recommendation G.711 (1972)~\cite{Cox_Etal:09:ITU-T-speech-coders,ITU-T_G.711}, which was widely deployed in public switched telephone network (PSTN) systems.  The G.711 encoder ingests 3.4~KHz of audio bandwidth digitized at   8~kilosamples per second (Ksps) with 13- or 14-bit amplitude quantization. It outputs 8-bit pulse code modulation (PCM) words using an A-law or $\mu$-law  quantizer for a  rate of 64~kilobits per second (Kbps).   This is a pure fixed-rate scalar quantization encoder; there is no frequency transformation nor entropy coding.  Speech coding is heavily constrained by the latency that humans can tolerate and application-specific processing, memory, and power  limitations.  Thus, entropy coding was not added to G.711 until 2009, when several  variable-length coding options, including Rice-Golomb coding, appeared in ITU-T Recommendation G.711.0~\cite{ITU-T_G.711.0}. 

A   feature that is becoming increasingly important as   high-fidelity media applications proliferate and speech   has to share bandwidth on multiplexed channels is Quality-of-Service (QoS) scalability.  ITU-T Recommendation G.711.1~\cite{ITU-T_G.711.1,HiwasakiOhmuro:09:ITU-T-G.711.1} is a backwards-compatible extension of G.711 that supports 16~Ksps   (7~KHz bandwidth) ``wideband'' speech while generating a layered (multiple bit rate) codestream that contains an embedded 8~Ksps G.711-compliant narrowband bitstream.  This is done  using a two-channel, 32-tap linear phase  \emph{pseudo quadrature mirror filter} (PQMF) bank to split the wideband input into 8~Ksps lowpass and highpass subbands.  Linear phase is desirable to avoid  nonlinear phase distortion in quantized speech. The quadrature mirror relation reduces  numerical filter bank design to the optimization of a single lowpass filter, and if that filter has linear phase then the overall analysis-synthesis transfer function will also have linear phase.  Unfortunately, the only two-channel FIR PR solutions  satisfying \emph{both} of these conditions are generalized Haar filter banks, so PQMF  banks like the one  in G.711.1   provide ``near-perfect,''  alias-free reconstruction with a linear phase transfer function that has \emph{approximately} constant  magnitude~\cite[Section~5.2]{Vaid93}.  

The G.711.1 lowband is encoded as  layer~0 using one of the core G.711 PCM algorithms.   An optional  enhancement bitstream (layer~1) encodes the residual  from layer~0 by adaptive allocation of anywhere from zero to three additional bits   per PCM codeword, constrained  to a rate of 16~Kbps, for an enhanced narrowband codestream with a   rate of 80~Kbps.  Wideband content is provided in layer~2 of  G.711.1  by coding the  highpass PQMF subband using a \emph{modified discrete cosine transform} (MDCT).  The $M$-channel MDCT can be regarded as a  windowed ``short-time discrete cosine transform''  in which the  signal is blocked into length-$2M$  blocks with 50\% overlap and tapered by a  window, much like the construction of short-time Fourier transforms.  A nontrivial fact is that, with proper window  and cosine transform design, one can save just $M$ output samples from each length-$2M$ block  and still have an invertible transformation.    Obtaining critical sampling (a  1:1 ratio of output to  input  samples) with 50\%  overlap to reduce blocking artifacts is clearly desirable in source coding applications, but it is far from obvious that one can do so while maintaining invertibility~\cite{Vaid93,StrNgu96,PainterSpanias:00:Perceptual-Coding}.

One can also interpret MDCTs  as  $M$-channel cosine-modulated filter banks  in which every filter  is a frequency-modulated version of a length-$2M$  ``prototype'' lowpass filter, or window.  Because   modulation is done with cosines rather than complex exponentials, a real-valued lowpass prototype yields real-valued bandpass filters.  This greatly reduces  design complexity: instead of  designing $M$  filters, one only needs  a single window satisfying appropriate conditions.  Moreover,  the polyphase representation can  be exploited to reduce implementation complexity by  decoupling   cosine modulation  from the lowpass prototype.  Thus, the filter bank can be implemented ``separably,'' using the $2M$-polyphase representation of the lowpass prototype and the $M$-point DCT-IV transform, which can be applied using FFT techniques~\cite{RaoYip90}.  This highlights an important development in source coding: the traditional distinction between filter bank-based ``subband coding'' and block-based ``transform coding'' has been  blurred by MDCTs since the lowpass filter is  usually applied using polyphase time-domain methods while the cosine modulation is performed using fast block transforms.

Layer~2 of the G.711.1 codestream is formed  using an 80-point MDCT with a Malvar sinusoidal window to split the 8~Ksps  PQMF highband into  $M=40$ frequency channels.  All but the four lowest-frequency channels are  quantized using interleave conjugate-structure vector quantization,   resulting in 80~bits per 5~ms frame, or 16~Kbps for layer~2.  The  bitrate for a 7~KHz ``wideband''  codestream containing  layers~0 and~2 is therefore 80~Kbps while the bitrate for  all three layers is 96~Kbps.    The layered structure of the G.711.1 wideband codestream enables a range of QoS scalability features, such as narrowband G.711-compliant  PSTN compatibility and partial mixing of teleconferencing  signals~\cite{HiwasakiOhmuro:09:ITU-T-G.711.1}.

Annex~D of G.711.1~\cite{ITU-T_G.711.1_AMD_4,MiaoLiuHuEtal:11:G.711.1-AnnexD} extends the G.711.1 layered codestream  to support ``superwideband'' (14~KHz) input  sampled at 32~Ksps with 16-bit precision.  The superwideband input is split into two 16~Ksps subbands  by a 32-tap linear phase PQMF bank  similar to the PQMF bank in the G.711.1 core encoder.  The  16~Ksps ``wideband'' lowpass subband is passed to the G.711.1 core encoder, where it is split again by the core  PQMF bank into lowband and highband signals and encoded as described above, giving Annex~D a two-level Mallat-style subband decomposition.  Two  highband enhancement layers   improve  performance in the 6.4--8.0~KHz range.  

A major  change for G.711.1  in  Annex~D is dynamic  classification of input into \emph{transient} and \emph{non-transient}  frames based on  the  Annex~D PQMF bank's 8--14~KHz  ``super higher band'' (SHB).  The SHB  is split into 80  channels by a 160-point  MDCT.  Its output is used  to classify non-transient frames into harmonic, normal, or noise-like frames and to switch between different  modes for quantizing SHB data using spectral and temporal envelope coding and MDCT-domain vector quantization.  Layering Annex~D  enhancement and SHB coding  on top of   G.711.1   codestreams creates four additional superwideband  modes with bitrates of 96--128~Kbps~\cite{ITU-T_G.711.1_AMD_4}.  The G.711.1 Annex~D  algorithm  was  developed jointly as a superwideband extension for ITU-T Recommendation G.722~\cite{ITU-T_G.722,ITU-T_G.722_AMD_1,MiaoLiuHuEtal:11:G.711.1-AnnexD}, the first ITU-T wideband voice  standard, which uses a linear phase PQMF bank to encode wideband input in a two-channel adaptive DPCM  scheme.

\subsubsection{G.729}\label{sec:Coding:Speech:G.729}
ITU-T Recommendation G.729.1~\cite{ITU-T_G.729.1,VargaProustTaddei:09:ITU-T-G.729.1} is a wideband  extension of the    G.729 narrowband standard~\cite{ITU-T_G.729}, which is widely used for voice-over-IP (VoIP) communications.  A 64-tap linear phase PQMF bank splits the 16~Ksps input into 8~Ksps lowpass and highpass subbands, as in G.711.1.  The lowband is encoded using code-excited linear prediction (CELP) to produce a  G.729-compatible core layer at 8~Kbps and one 4~Kbps narrowband enhancement layer.  Spectral and temporal envelope highband coding  at 2~Kbps yields a wideband codestream for  just 14~Kbps, while MDCT vector quantization creates 9 highband enhancement layers providing   scalability from 16 to 32~Kbps in 2~Kbps increments.  

Annex~E of G.729.1~\cite{ITU-T_G.729.1_AMD_6} supports superwideband input (14~KHz, 32~Ksps)  with 5  layers  providing rates from 36 to 64~Kbps.  Unlike the superwideband extensions for G.711.1 and G.722,  G.729.1 Annex~E does not  split the  input  using a  PQMF bank.  Instead, G.729.1 Annex~E employs essentially the same algorithm used to extend ITU-T Recommendation G.718~\cite{ITU-T_G.718,JelinekVaillancGibbs:09:G.718-new-embedded} for superwideband input~\cite[Annex~B]{ITU-T_G.718_AMD_2}.  The  32~Ksps input  is antialias-filtered using an IIR lowpass filter and subsampled to 16~Ksps for input to the G.729.1/G.718 core wideband encoders.  The full superwideband input  is simultaneously transformed  by a 640-channel MDCT  using a novel asymmetric window originally engineered for   G.718~\cite{ITU-T_G.718}.  The MDCT   7--14~KHz data is analyzed to classify individual frames as ``tonal'' or ``non-tonal'';  different MDCT vector quantization schemes  then create  layers extending the core wideband coding to superwideband.

\subsection{Audio Coding}\label{sec:Coding:Audio}
As speech coding covers wider bandwidths and  diverse multimedia content,  speech and general audio codecs are becoming more similar.  This is particularly true of their time-frequency analysis, which is driven largely by receiver characteristics; i.e., the human auditory system~\cite{www.cochlea.org}.  There are still  fundamental differences between  speech and general audio, however, such as  mature source models for speech,  limitations on  size, weight, and power  for mobile phones, and the desire for high-fidelity audio  to provide ``perceptually transparent''  coding across the entire auditory spectrum.  

The  lack of detailed source models for general audio  and the availability of greater computational power has driven high-fidelity audio  towards  adaptive quantization and  entropy coding based on  short-time quasi-stationary  modeling  of the  auditory system~\cite{RobinsonHawksford:99:Time-domain-auditory,PainterSpanias:00:Perceptual-Coding}.   The key feature of such  models  is the ``critical band'' theory of the cochlea as a spectrum analyzer  modeled by a nonuniform bank of  nonlinear (amplitude-dependent) bandpass filters exhibiting  psychoacoustic masking behavior.  This is approximated in practice by cascaded filter banks and MDCTs with quantization strategies that exploit perceptual masking of weak tones by nearby stronger tones.  

\subsubsection{MP3}\label{sec:Coding:Audio:MP3}
One of the earliest high-fidelity audio  standards to use perceptual modeling is  the MPEG-1 Part~3 standard (1992)~\cite{ISO_11172_3,Noll:97:MPEG-Digital-audio,PainterSpanias:00:Perceptual-Coding} created by the ISO/IEC Motion Picture Experts Group (MPEG).  MPEG-1 audio uses a uniform 32-channel, critically sampled, 512-point, near-perfect reconstruction cosine-modulated PQMF bank based on the DCT-III, reflecting the limitations of early-1990s filter bank technology.  Layers~1 and~2 use perceptual modeling to perform dynamic time-frequency bit allocation for  block companding.  
Layer~3 (the ``MP3'' algorithm) refines the frequency partition  by cascading the 32 uniform (750~Hz bandwidth) PQMF channels with adaptively switched MDCTs.  An 18-channel MDCT creates narrow frequency bands (41.67~Hz) to resolve low- and mid-frequency critical bandwidths for  perceptual coding of stationary frames while a 6-channel MDCT provides better temporal resolution for mitigating pre-echo artifacts caused by  transient attacks.  Transition windows  preserve invertibility when switching between 6- and 18-channel MDCTs. Layer~3 also uses run-length and Huffman entropy coding.

\subsubsection{AAC}\label{sec:Coding:Audio:AAC}
The MPEG-2  Part~3  standard~\cite{ISO_13818_3,Noll:97:MPEG-Digital-audio,PainterSpanias:00:Perceptual-Coding}  defines an embedded multichannel (``5.1'') surround-sound codestream that is backwards compatibility with MPEG-1 two-channel stereo decoding and supports lower sampling rates than MPEG-1.  MPEG-2  also has a more advanced, non-backwards-compatible audio codec known as  {Advanced Audio Coding} (AAC)~\cite{ISO_13818_7}.  The Low Complexity and Main profiles of MPEG-2 AAC eliminate the front-end PQMF bank in MPEG-1 in favor of a single MDCT whose window size switches between 2048 points for stationary content and 256 points for transients.  The Scalable Sample Rate profile has a 4-channel front-end QMF bank followed by MDCTs to enable an embedded codestream supporting multiple bitrates.   The MPEG-4 Part~3  standard~\cite{ISO_14496-3,Purnhagen:99:Overview-MPEG-4,PainterSpanias:00:Perceptual-Coding} includes  MPEG-2 AAC  in Subpart~4 for ``general audio''  coding and adds a great many other object-based audio coding tools to create a QoS-scalable toolkit supporting a vast range of audio modalities, including speech, parametric (model-based) audio, synthetic audio, MIDI, surround-sound, and various lossless   modes.  

Based on spectral analysis of the input, the AAC MDCT can  switch between a sinusoidal window  for narrowband selectivity and a Kaiser-Bessel window~\cite{Shlien:97:modulated-lapped-transform} for greater stopband attenuation.    The perceptually driven design of  Kaiser-Bessel MDCT windows  was pioneered for the Dolby AC-2 and AC-3 algorithms~\cite{Todd:94:AC-3,PainterSpanias:00:Perceptual-Coding};  AC-3 is the audio codec for the U.S.\ HDTV broadcast standard~\cite{ATSC_A52:2010}.  One novel feature of the AC-3 MDCT is that it can switch between 512-point and 256-point windows without using intervening transition windows to preserve invertibility~\cite{Shlien:97:modulated-lapped-transform}.

\subsection{Image Coding}\label{sec:Coding:Image}
Unlike audio, imagery essentially never contains sinusoidal waveforms, and the  audio coding strategy of transforming a source into hundreds of narrowband, quasi-stationary channels with long block lengths is  inappropriate for images.  A good first approximation for  natural images is to regard them as composed of  smooth but irregularly shaped regions separated by abrupt jump discontinuities that are readily discerned by the highpass  characteristic of the human visual system.  While fine textures commonly exist in continuous-tone images, preservation of fine texture generally is not as perceptually important as preservation of sharp edges between  regions.  Image transforms  thus need to provide highly localized (sparse) representations of high-frequency transients (edges), a requirement that does not match up well with the  properties of Fourier analysis.

\subsubsection{JPEG-1}\label{sec:Coding:Image:JPEG}
The most widely used international standard for  continuous-tone  imagery is the  standard produced by the ISO/IEC Joint Photographic Experts Group and known as JPEG (or JPEG-1)~\cite{ISO_10918_1,Wallace:91:JPEG,PenMit92}.  As a result of comprehensive engineering  and  perceptual studies in the 1984--88 time period~\cite{HudsonYasudaSebestye:88:international-standardisation,WallaceVivianPoulsen:88:Subjective-testing,LegerMitchellYamazaki:88:Still-picture}, the JPEG committee chose a block-transform algorithm using a two-dimensional nonoverlapping  $8\times 8$-pixel DCT-II~\cite{RaoYip90}.  The relatively small block size (larger blocks would have provided more coding gain) represents a compromise reflecting the need for good spatial localization of information in the transform domain. Moreover, the committee was sensitive to the risk of imposing high implementation costs (for the 1980s) in a first-generation communication  standard.

A bank of uniform scalar quantizers is applied to the DCT output, with relative bit allocation between different frequencies  given by a perceptually tuned quantization matrix and absolute  bit rate controlled by a single scalar parameter.  JPEG-1 offers  Huffman  coding as well as a higher-performance/higher-complexity binary arithmetic coding option.  The DCT architecture creates a limited amount of QoS scalability.  Progressive transmission across slow links can be provided by transmitting  DCT coefficients in order from lowest to highest spatial frequencies.  Hierarchical  scalability can be obtained by decoding and rendering an 8:1 reduced-resolution thumbnail of an image using only the DC coefficient from each $8\times 8$  block.  ``Reversibility'' (lossless coding) is possible by entropy encoding and transmitting the residual from a lossy  JPEG-1 representation.

\subsubsection{WSQ}\label{sec:Coding:Image:WSQ}
In the late 1980s the U.S.\ Federal Bureau of Investigation (FBI) decided to digitize the U.S.\ criminal  fingerprint database, which at the time consisted  of an acre of filing cabinets holding over 100~million   fingerprint cards.   They found that JPEG-1 blocking artifacts were unavoidable  at entropies below about 0.8~bits/pixel and interfered with both human and automated forensic end-users.  After working with researchers at Yale, Washington University, and Los Alamos National Lab, the FBI chose a 2-D DWT approach using cascaded two-channel linear phase  PR filter banks, optimal  subband rate allocation, uniform scalar quantization, and adaptive Huffman coding.  The resulting \emph{Wavelet/Scalar Quantization} (WSQ) specification~\cite{WSQ-v3.1:2010,BradBris94,BBOH96} included a  scheme for handling linear phase filter banks at image boundaries  by symmetrically extending and periodizing  finite-length input vectors, much like the interpretation of the DCT-II as the DFT of a symmetrically extended vector~\cite{Bris96b,Bris98}.

\subsubsection{JPEG~2000}\label{sec:Coding:Image:JPEG2000}
By the mid-1990s it was clear that   subband  coding  offered significant improvements in both rate-distortion performance and QoS scalability over   JPEG-1, so the  ISO committee created a new work item known as JPEG~2000 to address the growing list of applications inadequately served by  JPEG-1~\cite{ISO_15444_1,TaubMarc02,AcharyaTsai:04:JPEG2000-Standard,Lee:05:J2K-Retrospective}.  JPEG~2000 was heavily influenced by the highly scalable embedded subband coding approach  in the PhD dissertation of Taubman~\cite{Taubman:94:PhD-thesis}.  The theory of lifting factorizations~\cite{Sweldens96,Sweldens:98:SIAM-lifting-scheme,DaubSwel98} also had a big impact on JPEG~2000, which uses lifting  to specify  implementation and signaling of PR filter banks.  The ability to implement filter banks with dyadic  lifting coefficients as nonlinear integer-to-integer (``reversible'') transforms~\cite{CaldDaubSwelYeo:ACHA-98:ints-to-ints} is exploited to provide  lossy-to-lossless QoS scalability, greatly improving on the  lossless coding features of JPEG-1.  JPEG~2000 also uses symmetric extension boundary handling, which can be implemented directly in terms of lifting factorizations~\cite{Bris07d,WohlBris08}.  

JPEG~2000 Part~1 has one irreversible  filter bank (the same  one used in WSQ) and one reversible option~\cite[Annex~F]{ISO_15444_1}.  Both  are WS PR wavelet filter banks suitable for  cascaded DWT decompositions.  JPEG~2000 Part~2, \emph{Extensions}, allows users to signal  user-defined WS PR filter banks~\cite[Annex~G]{ISO_15444_2} or arbitrary  two-channel PR filter banks (including HS and paraunitary filter banks)~\cite[Annex~H]{ISO_15444_2}.    Part~2 and Part~10, \emph{Extensions for three-dimensional data}~\cite{ISO_15444_10}, include algorithms for using filter banks to decorrelate multi-banded images such as  multispectral or volumetric image cubes.  JPEG~2000 Part~9, \emph{Interactivity tools, APIs, and protocols} (JPIP)~\cite{ISO_15444_9}, exploits the joint space-frequency localization of DWT decompositions and the bit-plane localization of JPEG~2000's binary arithmetic coding  to provide a client-server protocol enabling highly scalable interactive retrieval of compressed data.

\subsubsection{NITFS}\label{sec:Coding:Image:NITFS}
JPEG~2000 Part~1  is used in the U.S. National Imagery Transmission Format Standard (NITFS)~\cite{NITFS-2.1} for conventional military imagery, and a JPIP profile  is provided in~\cite{MISB_RP0811.0:2009}. Much work remains to be done on applying JPEG~2000 Part~2 extensions to the many unconventional  modalities that arise in military applications, such as multi- and hyperspectral imagery, infrared, SAR, LIDAR, etc.  One military application that has received attention is  \emph{large volume streaming data} (LVSD)~\cite{MISB_RP0705.2:2008}, which consists of wide-area surveillance video often collected from airborne platforms.  Although LVSD is video imagery, the LVSD profile uses intraframe (non-motion-compensated) JPEG~2000 Part~1 coding.   LVSD applications are characterized by single-frame image analysis requirements, very large frame sizes (up to a gigapixel or more), slow frame rates (often less than 10~frames/sec.), and, sometimes, high bit depths or unconventional modalities, all of which weigh against MPEG solutions.  Another benefit of  JPEG~2000 in LVSD applications is the  JPIP profile~\cite{MISB_RP0811.0:2009}, which facilitates single-frame analysis of gigapixel imagery.

\subsubsection{DCI}\label{sec:Coding:Image:DCI}
Another video application that has adopted JPEG~2000 intraframe coding in preference to motion-compensated  coding is the {Digital Cinema Initiatives} (DCI) specification for theater distribution of  feature films~\cite{DCI_DCSS_1.2:2008}.  For the DCI application, having a resolution-scalable format that supports extremely high fidelity (``better than traditional 35mm prints'') is more important than meeting  stringent bandwidth and hardware complexity constraints of the sort MPEG standards are designed to satisfy.  E.g., the maximum allowable DCI bit rate is 250~Mbps for the video signal (not including audio) whereas the maximum allowable video  rate in the Blu-Ray format (which  uses MPEG-4 and  VC-1) is 40~Mbps.

\subsection{Video Coding}\label{sec:Coding:Video}
DWTs have yet to achieve commercial success in motion-compensated video coding.  Video standards from MPEG-1 up through MPEG-4/H.264 \emph{Advanced Video Coding} (AVC)~\cite{ISO_14496-10,ITU-T_H.264,WiegandSullivanBjontegLuthra:03:H.264-AVC,PuriChenLuthra:04:Video-coding-H.264} and  the new \emph{High-Efficiency Video Coding} (HEVC) standard~\cite{ISO_23008_HEVC_DIS_8} have consistently used nonoverlapping block DCTs to code motion  prediction residuals.  In  AVC the $16\times 16$-pixel motion prediction macroblocks may be partitioned into sub-macroblocks as small as $4\times 4$ to improve  spatial localization, so AVC employs an integer-to-integer approximation of the $4\times 4$ DCT-II to  preserve the segmentation induced by  block motion compensation.  Blocking artifacts are ameliorated by deblocking filters within the motion compensation prediction loop.  Intra-frame coding in AVC uses a variety of  spatial prediction filters to reduce spatial redundancy, so the DCT block transform encoder is compressing prediction residuals even in intra-coded frames.  As with MDCTs on audio residuals,   DCT-IIs provide  good transform coding performance on closed-loop  video prediction residuals. 

When the  Motion Picture Experts Group called for proposals for a  scalable extension to MPEG-4/H.264 AVC,  closed-loop motion compensation posed many challenges to both temporal and spatial  scalability, and open-loop motion-compensated 3-D subband coding was expected to offer competitive alternatives; see Ohm~\cite{Ohm:05:Advances-Scalable-Video} for an exposition of the situation circa 2005.  Twelve of the 14 proposals submitted involved some form of  3-D discrete wavelet transform, but none of the wavelet  proposals was able to overcome the considerable head start enjoyed by the AVC approach.  In the end, the ISO/IEC/ITU-T Joint Video Team used AVC's reference picture memory control  to enable hierarchical closed-loop temporal prediction and formed a layered codestream  with inter-layer prediction to enable spatial resolution scalability~\cite{SchwarzMarpeWiegand:07:Overview-scalable}.  The  extension was approved as Amendment~3 to MPEG-4/H.264 AVC and incorporated  as~\cite[Annex~G: Scalable Video Coding]{ISO_14496-10}.    It is an open question whether promising wavelet transform approaches such as~\cite{SeckerTaubman:03:LIMAT} will eventually gain a foothold in the motion-compensated video coding market.

\section{Review of Previous Results}\label{sec:Review}
\subsection{Group Lifting Structures}\label{sec:Review:GLS}
In previous work \cite{Bris:10:GLS-I,Bris:10b:GLS-II} the author developed a theory of \emph{group lifting structures} that provides a group-theoretic framework for parameterizing classes of  filter banks of practical interest, including linear phase FIR  filter banks.
A major impetus for using group theory to describe filter banks is the fact that PR  filter banks do not naturally form  vector spaces but \emph{do} form  matrix groups in the polyphase representation (Figure~\ref{DS_poly}). With few exceptions (e.g.,~\cite{Park:04:Symbolic-computation-signal,LebrunSelesnic:04:Grobner-bases-wavelet}) most research to date on multirate filter banks has relied almost exclusively on mathematical tools from linear algebra and polynomial factorization (notably the Euclidean algorithm).

The  paper \cite{Bris:10:GLS-I} introduced group lifting structures and constructed examples for the two principal classes of linear phase filter banks~\cite{NguVaid:89:2-channel-PR-LP}, the WS and HS classes exemplified in Figure~\ref{WS-HS}.  A uniqueness theorem \cite[Theorem~1] {Bris:10:GLS-I} was proven for ``irreducible'' lifting factorizations generated by a group lifting structure that satisfies suitable hypotheses.  The second paper \cite{Bris:10b:GLS-II} showed that the WS and HS group lifting structures satisfy the hypotheses of the  uniqueness theorem and therefore have irreducible group lifting factorizations that are either unique (in the WS case) or ``unique modulo rescaling'' (in the HS case).  These unique factorization results are significant because elementary matrix decompositions, including lifting factorizations, are  highly nonunique in general.

The original motivation for  group lifting factorizations arose from the author's work on the ISO/IEC JPEG~2000 image coding standards~\cite{ISO_15444_1,ISO_15444_2}, which are based on subband coding using  DWTs (or ``wavelet transformations'' in the language of~\cite{ISO_15444_1,ISO_15444_2}).  In particular, \cite[Annex~G]{ISO_15444_2} is devoted  to the  signaling and lifting implementation of WS filter banks.    One consequence of the uniqueness theorem for WS group lifting factorizations \cite[Theorem~1]{Bris:10b:GLS-II} is that a WS filter bank can be specified in JPEG~2000 Annex~G-compliant syntax in one and only one way \cite[Corollary~1]{Bris:10b:GLS-II}.  Another consequence was disproving an assertion made in~\cite[p.~294]{TaubMarc02} that WS filter banks whose filters differ in length by  two (the mimimal amount  for this class) always have lifting factorizations using first-order HS (type~II) linear phase lifting filters.   A consequence of the uniqueness theorem for HS group lifting factorizations \cite[Theorem~2]{Bris:10b:GLS-II} is that there exist many HS filter banks, including the example filter banks in Annexes~H.4.1.2.1 and H.4.1.2.2 of~\cite{ISO_15444_2},  that \emph{cannot} be lifted from the Haar filter bank,
\begin{equation}\label{Haar}
H_0(z) = 0.5(z+1),\quad H_1(z) = -z+1,
\end{equation}
using  \emph{whole-sample antisymmetric} (WA, or type~III) linear phase lifting filters.

Further motivation for  group lifting factorizations  is provided by the  present paper, which shows that the theory developed in \cite{Bris:10:GLS-I,Bris:10b:GLS-II} allows one to characterize both classes of linear phase filter banks  in group-theoretic terms.  This means  we can describe the structure of the  WS class, whose polyphase matrices form a  group, in terms of standard group-theoretic constructs  involving the building blocks of lifting factorization:  upper and lower triangular lifting matrix groups and a group of diagonal gain-scaling matrices.  Abstract algebra has provided valuable  tools in other branches of  signal processing, notably the application  of finite (Galois) fields to channel coding, and the author hopes that the group-theoretic perspective will  provide useful and practical insights into subband coding.

\subsection{Notation and Terminology}\label{sec:Review:Notation}
``$X\equiv \cdots$'' means that $X$ is equal to $\cdots$ \emph{by definition}.  Column vectors are denoted in bold math italic while matrices are in bold upright fonts, e.g., $\mathbf{A}\bsy{x}=\bsy{b}.$  Algebraic groups are denoted in calligraphic fonts; $\mathscr{G}<\mathscr{H}$ means  $\mathscr{G}$ is a subgroup of $\mathscr{H}$ while $\mathscr{G} \vartriangleleft\mathscr{H}$ means  $\mathscr{G}$ is a \emph{normal} subgroup.  If $X\subset\mathscr{H}$ then the subgroup of $\mathscr{H}$ generated by $X$ is
\[ \langle X\rangle\equiv \{x_1\cdots x_n\colon x_i\in X\;\mbox{or}\; x_i^{-1}\in X\} < \mathscr{H}.  \]
$\mathscr{G}\cong\mathscr{H}$ means  $\mathscr{G}$ and $\mathscr{H}$ are isomorphic.  $\Aut(\mathscr{G})$ is the group of  automorphisms of $\mathscr{G}$.  The digit 1 denotes  various group identity elements, identity transformations, and  trivial groups.  The identity matrix, however, is denoted $\mathbf{I}$, as usual.

\subsubsection{The polyphase-with-advance representation}\label{sec:Intro:Notation:Polyphase}
\begin{figure}[tb]
  \begin{center}
    \includegraphics{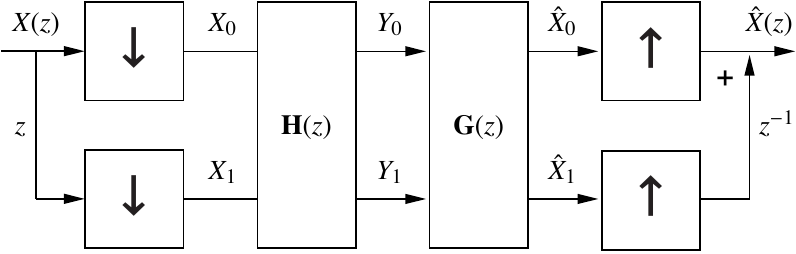}
    \caption{The polyphase-with-advance filter bank representation.}
    \label{DS_poly}
  \end{center}
\end{figure}
Figure~\ref{DS_poly} depicts the polyphase-with-advance representation \cite{VettKov95,StrNgu96,DaubSwel98,BrisWohl06} of a two-channel {FIR} PR filter bank~\cite{Vaid93}.  All polyphase matrices $\mathbf{H}(z)$ studied in this paper will be  polyphase-with-advance analysis matrices.  {FIR} polyphase matrices with {FIR} inverses are characterized by:
\begin{equation}\label{FIR_PR}
|\mathbf{H}(z)| \equiv \det\mathbf{H}(z)  = \check{a}z^{-\dd};\; \check{a}\neq 0,\ \dd\in\mathbb{Z}.
\end{equation}
As noted in~\cite{BrisWohl06,Bris:10:GLS-I}, the family $\mathscr{F}$ of all such FIR PR filter banks forms a nonabelian (i.e., noncommutative) group under matrix multiplication called the \emph{FIR filter bank group}.  The \emph{unimodular group}, $\mathscr{N}$, is the normal subgroup of  $\mathscr{F}$ consisting of all matrices of determinant~1, 
\begin{equation}\label{DS_FIR_PR}
|\mathbf{H}(z)| = 1\,.
\end{equation}

Daubechies and Sweldens~\cite{DaubSwel98} proved that every unimodular FIR matrix has a {\em lifting factorization} (or \emph{lifting cascade}),
\begin{equation}\label{anal_lift_cascade}
\mathbf{H}(z) = \mathbf{D}_K\,\mathbf{S}_{N-1}(z)\cdots\mathbf{S}_1(z)\,\mathbf{S}_0(z).
\end{equation}
We shall work with slightly more general lifting decompositions in which  $\mathbf{H}(z)$ is lifted from a \emph{base} filter bank, $\mathbf{B}(z)$:
\begin{equation}\label{lift_from_B}
\mathbf{H}(z) = \mathbf{D}_K\,\mathbf{S}_{N-1}(z)\cdots\mathbf{S}_0(z)\,\mathbf{B}(z).
\end{equation}
The \emph{gain-scaling matrix}, $\mathbf{D}_K$, is a  unimodular diagonal constant matrix with  \emph{scaling factor} $K\neq 0$,
\begin{equation}\label{gain_matrix}
\mathbf{D}_K \equiv \diag(1/K,\,K).
\end{equation}
The \emph{lifting matrices}, $\mathbf{S}_i(z)$, are unimodular upper or lower triangular matrices with ones on the diagonal and  \emph{lifting filters}, $S_i(z)$, on the off-diagonal, given via the homomorphisms
\begin{equation}\label{monomorphisms}
\upsilon(S(z)) \equiv
        \left[ \begin{array}{cc}
                1  & S(z) \\
                0  & 1
        \end{array}\right],\quad
\lambda(S(z)) \equiv
        \left[ \begin{array}{cc}
                1  & 0\\
                S(z)  & 1
        \end{array}\right].
\end{equation}

The \emph{update characteristic}~\cite[Annex~G.1]{ISO_15444_2} of a lifting step is a binary flag, $m$, indicating whether the lift is a lowpass update ($m=0$; upper triangular  matrix) or a highpass update ($m=1$; lower triangular  matrix).   A  lifting cascade is \emph{irreducible}~\cite[Definition~3]{Bris:10:GLS-I} if the  lifting matrices are nontrivial ($\mathbf{S}_i(z)\neq \mathbf{I}$) and strictly alternate between lower and upper triangular.  As noted in~\cite{Bris:10:GLS-I}, any lifting cascade can be simplified to irreducible form using matrix multiplication.

\subsubsection{Linear phase {FIR} {PR} filter banks}\label{sec:Intro:Notation:Linear}
In~\cite{BrisWohl06,Bris:10:GLS-I} unimodular WS and HS filter banks were normalized to satisfy \emph{delay-minimized} conventions.  Specifically, unimodular WS filter banks are normalized so that the group delay of $H_0(z)$ is $d_0=0$ and the group delay of $H_1(z)$ is $d_1=-1$.
This is equivalent to having  $\mathbf{H}(z)$ satisfy the intertwining relation
\begin{equation}\label{WS_intertwining}
\mathbf{H}(z^{-1}) = \bsy{\Lambda}(z)\mathbf{H}(z)\bsy{\Lambda}(z^{-1}),\quad \bsy{\Lambda}(z)\equiv\diag(1,\,z^{-1}).
\end{equation}

Unimodular filter banks satisfying~(\ref{WS_intertwining}) form a subgroup of $\mathscr{N}$ called the \emph{unimodular WS  group}, $\mathscr{W}$~\cite[Definition~8]{Bris:10:GLS-I}.  It was shown in~\cite{Bris:10:GLS-I} that the  lowpass lifting updates satisfying~(\ref{WS_intertwining}) have lifting filters $S_i(z)$ that are \emph{half}-sample symmetric about $1/2$ and belong to the additive group of Laurent polynomials
\begin{equation}\label{lowpass_HS}
\mathscr{P}_0 \equiv \left\{S(z)\in\mathbb{R}[z,z^{-1}] \colon S(z^{-1})=zS(z)\right\}. 
\end{equation}
Each  filter in $\mathscr{P}_0$ is mapped isomorphically to a corresponding upper triangular lifting matrix $\upsilon(S(z))$ in a multiplicative but abelian (commutative) matrix group $\mathscr{U}\equiv\upsilon(\mathscr{P}_0)$, making $\mathscr{U}\cong\mathscr{P}_0$~\cite[Section~III-A]{Bris:10:GLS-I}.  Similarly, lower triangular lifting matrices satisfying (\ref{WS_intertwining}) have lifting filters  that are half-sample symmetric about $-1/2$ and belong to 
\begin{equation}\label{highpass_HS}
\mathscr{P}_1  \equiv \left\{S(z)\in\mathbb{R}[z,z^{-1}]  \colon S(z^{-1})=z^{-1}S(z)\right\}. 
\end{equation}
The function  $\lambda$   maps  $\mathscr{P}_1$  isomorphically onto an abelian group $\mathscr{L}\equiv\lambda(\mathscr{P}_1)$ of lower triangular lifting matrices.

HS filter banks are normalized so that {both} group delays are $d_0=d_1=-1/2$.  This is equivalent to having  $\mathbf{H}(z)$ satisfy
\begin{equation}\label{HS_anal_mirror_DS}
\mathbf{H}(z^{-1}) = \mathbf{L}\mathbf{H}(z)\mathbf{J},
\end{equation}
where
\begin{equation}\label{LandJ}
\mathbf{J}\equiv
        \left[ \begin{array}{cc}
                0  & 1\\
                1  & 0
        \end{array}\right], \quad
\mathbf{L} \equiv
        \left[ \begin{array}{cc}
                1  & 0\\
                0  & -1
        \end{array}\right].
\end{equation}
It was shown in~\cite{BrisWohl06} that delay-minimized HS polyphase matrices do \emph{not} form a group.  The \emph{unimodular HS class}, $\mathfrak{H}$ \cite[Definition~9]{Bris:10:GLS-I}, is the set of all unimodular HS filter banks  satisfying (\ref{HS_anal_mirror_DS}).  HS filter banks with \emph{unequal}-length filters $H_0(z)$ and $H_1(z)$ can be lifted from \emph{equal}-length HS ``base'' filter banks using lifting matrices with  WA lifting filters.  Equal-length HS base filter banks can in turn be factored into (non-WA)  lifting steps using the general  machinery of~\cite{DaubSwel98}.   WA lifting filters belong to the additive ``antisymmetric'' group
\[  \mathscr{P}_a \equiv \left\{S(z)\in\mathbb{R}[z,z^{-1}] \colon S(z^{-1})=-S(z)\right\}.  \]
The upper and lower triangular WA lifting matrix groups are
\[  \mathscr{U}\equiv\upsilon(\mathscr{P}_a),\quad \mathscr{L}\equiv\lambda(\mathscr{P}_a). \]

\subsubsection{Group lifting structures}\label{sec:Intro:Notation:Group}
A  \emph{group lifting structure}~\cite[Definition~6]{Bris:10:GLS-I} is an ordered four-tuple, 
\begin{equation}\label{group_lifting_structure}
\mathfrak{S}\equiv(\mathscr{D},\mathscr{U},\mathscr{L},\mathfrak{B}).
\end{equation}
The abelian  group $\mathscr{D}\equiv \{\mathbf{D}_K\colon K\in\mathscr{R}\}$ consists  of gain-scaling matrices~(\ref{gain_matrix}) parameterized by a multiplicative group,
\begin{equation}\label{D_iso}
\mathscr{R}<\mathbb{R}^*\equiv \mathbb{R} \backslash \{0\},\quad
\mathbf{D}\colon \mathscr{R}\stackrel{\cong}{\longrightarrow}  \mathscr{D}.
\end{equation}
$\mathscr{U}$ and $\mathscr{L}$ are abelian groups of upper and lower triangular lifting matrices, and $\mathfrak{B}$ is a set  of base filter banks.  

The \emph{lifting cascade group}, $\mathscr C$, generated by $\mathfrak{S}$ is the nonabelian  subgroup of $\mathscr N$ generated by $\mathscr{U}$ and $\mathscr{L}$ \cite[Definition~6]{Bris:10:GLS-I},
\begin{equation}\label{lifting_cascade_group}
\mathscr{C} \equiv \langle\mathscr{U\cup L}\rangle =
\left\{\mathbf{S}_N\cdots\mathbf{S}_1 \colon N\geq 1,\;\mathbf{S}_i\in\mathscr{U\cup L}\right\}. 
\end{equation}
The \emph{scaled lifting group}, $\mathscr{S}$, generated by $\mathfrak{S}$ is the subgroup of $\mathscr{N}$ generated by $\mathscr{D}$ and $\mathscr{C}$,
\begin{equation}\label{scaled_lifting_group}
\mathscr{S}\equiv\langle\mathscr{D\cup C}\rangle\,.
\end{equation}
The universe  of \emph{all} filter banks generated  by $\mathfrak{S}$ is 
\[ 
\mathscr{DC}\mathfrak{B} \equiv 
\left\{\rule[-1pt]{0pt}{12pt}\mathbf{DCB} \colon
\mathbf{D}\in\mathscr{D},\;\mathbf{C}\in\mathscr{C}\,,\;\mathbf{B}\in\mathfrak{B}\right\}.
\]

A gain-scaling  matrix $\mathbf{D}_K\in\mathscr{D}$ acts on polyphase matrices via the \emph{inner automorphism} $\gam{K}\equiv\gamma_{\ssst\mathbf{D}_{\!K}}$,
\begin{equation}\label{conjugation_operator}
\gam{K}\mathbf{E}(z) \equiv \mathbf{D}_K\,\mathbf{E}(z)\,\mathbf{D}_{K}^{-1} .
\end{equation}
We use $\gamma$ to denote the homomorphism 
\begin{equation}\label{defn_gamma}
\gamma\colon\mathscr{D}\rightarrow\Aut(\mathscr{N}). 
\end{equation}
A group $\mathscr{G}$ of polyphase matrices is called \emph{$\mathscr{D}$-invariant} if $\mathscr{D}$ normalizes $\mathscr{G}$, i.e., $\gam{K}\mathscr{G} = \mathscr{G}$ for all $\mathbf{D}_K\in\mathscr{D}$, in which case we may regard $\gamma$ as a homomorphism of $\mathscr{D}$ into $\Aut(\mathscr{G})$.   A group lifting structure is called $\mathscr{D}$-invariant if $\mathscr{U}$ and $\mathscr{L}$, and therefore $\mathscr{C}$, are $\mathscr{D}$-invariant groups.

The lifting cascade~(\ref{lift_from_B}) is called \emph{strictly polyphase order-increasing} (or just \emph{order-increasing}) if the polyphase orders of the partial products for $0\leq n<N$, $\mathbf{S}_{-1}(z)\equiv\mathbf{I}$,  satisfy
\[ 
\order\left(\mathbf{S}_{n}(z)\cdots\mathbf{B}(z)\right) > \order\left(\mathbf{S}_{n-1}(z)\cdots\mathbf{B}(z)\right).
\]
A group lifting structure is  called order-increasing if every irreducible cascade in 
\mbox{$\mathscr{C}\mathfrak{B}$} is order-increasing.  It is a non-trivial fact that the linear phase group lifting structures for WS and HS filter banks are order-increasing~\cite{Bris:10b:GLS-II}.   

If $\mathfrak{S}$ is a $\mathscr{D}$-invariant, order-increasing group lifting structure,
the unique factorization theorem~\cite[Theorem~1]{Bris:10:GLS-I} says that all irreducible group lifting factorizations of  $\mathbf{H}(z)\in \mathscr{DC}\mathfrak{B}$ are ``equivalent modulo rescaling.'' Specifically, given two irreducible factorizations in $\mathscr{DC}\mathfrak{B}$ of the same matrix,
\begin{eqnarray*}
\mathbf{H}(z) & = & \mathbf{D}_K\,\mathbf{S}_{N-1}(z) \cdots\mathbf{S}_0(z)\, \mathbf{B}(z)  \\
        & = & \mathbf{D}_{K'}\,\mathbf{S}'_{N'-1}(z)\cdots \mathbf{S}'_0(z)\,\mathbf{B}'(z),
\end{eqnarray*}
the theorem states that the number of lifting steps is the same, $N'=N$, with base filter banks related by gain rescaling
\begin{equation}\label{defn_alpha} 
\mathbf{B}'(z) = \mathbf{D}_{\alpha}\,\mathbf{B}(z),\quad
\alpha\equiv K/K',
\end{equation}
and lifting steps related by inner automorphisms,
\begin{equation}\label{almost_unique_factors} 
\mathbf{S}'_i(z) = \gamma_{\!\alpha}\mathbf{S}_i(z)\,,\quad\mbox{$i=0,\ldots,N-1$.}
\end{equation}
We express this by saying that irreducible lifting factorizations in $\mathfrak{S}$ are ``unique modulo rescaling''~\cite[Definition~11]{Bris:10:GLS-I}.   This conclusion can be strengthened if, e.g., $\mathfrak{B}=\{\mathbf{I}\}$ (as with the  WS group, $\mathscr{W}$), in which case the only possibility is $\alpha=1$  and we obtain  \emph{unique} irreducible group lifting factorizations.

\section{Free Product Structure of the Lifting Cascade Group}\label{sec:Free}
We begin our study of  lifting cascade groups by reviewing the definitions and  properties of free groups and free products of groups.  Of particular importance is the definition of free products in terms of a \emph{universal mapping property} that  provides the key to the proof of our main result in Section~\ref{sec:Free:Cascade}.

\subsection{Free Groups}\label{sec:Free:Groups}
``Free'' groups are generated by ``relation-free'' generators, a notion familiar from linear algebra where the relation-free property is called \emph{linear independence} and a set of linearly independent generators for a vector space is called a \emph{basis}.  In contrast to the situation in linear algebra, in which \emph{every} vector space has a basis, a group with a set of relation-free generators is a rather special object, called a \emph{free group}.

\subsubsection{Defining free groups}\label{sec:Free:Groups:Define}
Rather than formally defining free groups in terms of relation-free generators, the algebra literature~\cite{Hungerford74,Rotman95,Robinson96} defines free groups in terms of a universal mapping (existence/uniqueness) property involving group homomorphisms.  This is an analogue of a standard result from linear algebra \cite[Theorem~IV.2.1]{Hungerford74}, where homomorphisms are known as \emph{linear transformations}.
\begin{prop}\label{prop:FreeVS}
Let $V$ be a vector space over a field $\mathbb{F}$ with an indexed subset 
\[ B=\jmath(I)=\{\bsy{b}_i:i\in I\}\subset V, \]
where $\bsy{b}_i=\jmath(i)$ for some index set $I$ and indexing function \mbox{$\jmath:I\rightarrow V.$}  The set $B$ is a basis for $V$ if and only if, for every $\mathbb{F}$-vector space $W$ and every function \mbox{$f:I\rightarrow W,$} there exists a unique linear transformation 
\[ \mathbf{T}:V\rightarrow W \]
such that $\mathbf {T}\bsy{b}_i=f(i)$ for all $i\in I$.
\end{prop}

\begin{figure}[tb]
  \begin{center}
    \includegraphics{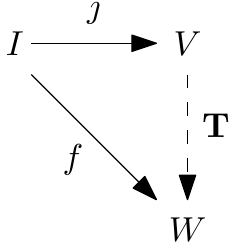}
    \caption{Commutative diagram for the universal mapping property characterizing a vector space with a basis indexed by the index set~$I$.}
    \label{fig:FreeVS}
  \end{center}
\end{figure}

Given $\jmath:I\rightarrow V$, the universal mapping property can be expressed in graphical terms by saying  there exists a unique linear transform $\mathbf T$, depending on $W$ and $f$, such that the diagram in Figure~\ref{fig:FreeVS}  commutes.  How do we interpret this universal mapping property in terms of more familiar linear-algebraic concepts?   When  $B$ is a basis for $V$, the linear transform $\mathbf T$ in Proposition~\ref{prop:FreeVS} is just the linear extension (to all of $V$) of the mapping that carries each basis vector $\bsy{b}_i\in B$ to the given vector $f(i)\in W$.   Uniqueness of $\mathbf T$ means that every linear transformation of $V$ is uniquely determined by its behavior on the basis  $B$.  According to Proposition~\ref{prop:FreeVS}, this universal mapping property  is actually  \emph{equivalent} to the statement that $B$ forms a basis for $V$.

Characterizing a  basis in terms of a commutative diagram like Figure~\ref{fig:FreeVS} is a ``categorical'' approach to the notion of relation-free generators that depends only on universal mapping properties.  Since this characterization makes no mention of properties specific to  vector spaces (e.g., linear independence), it can be directly generalized to other categories, such as the category of groups, by replacing linear transformations with homomorphisms of the appropriate type. 

\begin{figure}[tb]
  \begin{center}
    \includegraphics{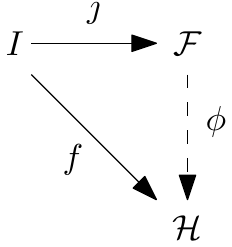}
    \caption{Commutative diagram for the universal mapping property defining a free group on the set $I$.}
    \label{fig:FreeGroups}
  \end{center}
\end{figure}

\begin{defn}[Free groups~\cite{Hungerford74,Rotman95,Robinson96}]\label{defn:FreeGroups}
Let $\mathscr{F}$ be a group with an indexed subset 
\[ \jmath(I)=\{g_i:i\in I\}\subset \mathscr{F}, \]
where $g_i=\jmath(i)$ for some index set $I$ and indexing function \mbox{$\jmath:I\rightarrow \mathscr{F}.$}  The group $\mathscr{F}$ is called a \emph{free group on the set $I$}  if and only if, for every group $\mathscr{H}$ and every function \mbox{$f:I\rightarrow \mathscr{H},$} there exists a unique group homomorphism
\[ \phi:\mathscr{F}\rightarrow \mathscr{H}  \]
such that $\phi\circ \jmath=f$;  i.e., such that $\phi(g_i)=f(i)$ for all $i\in I$.  This  is equivalent to saying  there exists a unique homomorphism $\phi$, depending on $\mathscr{H}$ and $f$, such that the diagram in Figure~\ref{fig:FreeGroups} commutes.
\end{defn}

\rem  If $\mathscr{F}$ is  free on  $I$ then  $\jmath$ must be  injective, and it also follows~\cite[Corollary~11.5]{Rotman95} that $\jmath(I)$ generates $\mathscr{F}$.    Abstract nonsense involving manipulations of commutative diagrams~\cite[Theorem~I.7.8]{Hungerford74} shows that $\mathscr{F}$ is determined up to isomorphism by the cardinality of $I$, denoted $|I|$. 

\begin{prop}\label{prop:FreeGroupUniqueness}
Let $\mathscr{F}$ be  free on $I$ and  $\mathscr{F'}$   free on $I'$. If $I$ and $I'$ have the same cardinality ($|I| = |I'|$) then  $\mathscr{F}$ and $\mathscr{F'}$ are isomorphic:
$\mathscr{F} \cong \mathscr{F'}.$
\end{prop}

The converse is also true~\cite{Rotman95,Robinson96}, and in light  of Proposition~\ref{prop:FreeGroupUniqueness} we sometimes speak of \emph{the} free group on $I$ or on $|I|$-many generators.
As mentioned above, the standard argument from linear algebra  showing that every vector space has a basis \emph{fails} for groups because a maximal set of relation-free elements need not generate the group.  For instance, a finite group is never free since \emph{all} elements satisfy a relation of the form $g^n = 1$ (the group identity element).  Free groups do exist, however:  infinite cyclic  groups, denoted 
\mbox{$\langle x \rangle \equiv \{x^n\colon n\in\mathbb{Z}\}$} 
in multiplicative notation, are infinite groups on one generator, $x$.  Such groups are free with $|I|=1$ and are isomorphic to the free  \emph{additive} group $\mathbb{Z}$.

Free groups are ``universal groups'' in the following sense.  Suppose  $\mathscr{H}$ is \emph{any} group, and let $X$ be a subset of generators indexed by a set $I$:
\[  \mathscr{H}=\langle X \rangle \quad \mbox{where}\quad X = \{h_i:i\in I\}\subset \mathscr{H}. \]
Let $\mathscr{F}$ be a free group on $I$ and define $f(i)\equiv h_i$ for each $i\in I$.  Since $\mathscr{F}$ is free there exists a unique homomorphism 
\mbox{$\phi:\mathscr{F} \rightarrow \mathscr{H}$}
such that Figure~\ref{fig:FreeGroups} commutes, and $\phi$ maps $\mathscr{F}$ \emph{onto} $\mathscr{H}$ because the $h_i$ generate $\mathscr{H}$.  It follows~\cite[Corollary~I.9.3]{Hungerford74}, \cite[Corollary~11.2]{Rotman95} that $\mathscr{H}$ is isomorphic to a quotient  of $\mathscr{F}$,
\[ \mathscr{H}\cong \mathscr{F}/\ker\phi. \]
This of course begs the question of whether, given an index set $I$, there always exists a free group  on $I$.

\subsubsection{Constructing free groups}\label{sec:Free:Groups:Construct}
There is a constructive  procedure, the ``reduced word construction,'' that generates a canonical free group on any given index set, $I$, and therefore (by Proposition~\ref{prop:FreeGroupUniqueness}) generates all free groups up to isomorphism.  The reduced word construction and its generalization  to free products  inspired one of the main results of this paper, Theorem~\ref{thm:Cascade}.   A rigorous treatment of the reduced word construction is more technical than we indicate in the following  (see~\cite{Hungerford74,Rotman95}), so the proof of Theorem~\ref{thm:Cascade}  avoids these technical details  by using a universal mapping characterization of free products.  The intuition behind the theorem, however, stems directly from the reduced word construction.  Before tackling free products, we first outline the  construction of free groups.

Given  a set $I$,  create an alphabet $X$ containing two distinct formal tokens, denoted $x_i$ and  $x_i^{-1}$, for  each $i\in I$.  A \emph{word}, $w$, on this token alphabet is a finite string of tokens,
\[ w = t_1t_2\ldots t_n, \]
where each $t_k$ equals some $x_i$ or  $x_i^{-1}$.  The \emph{inverse} of the above word, denoted $w^{-1}$, is defined to be
\[ w^{-1} \equiv t_n^{-1}\ldots t_2^{-1}t_1^{-1}, \]
where $(x_i^{-1})^{-1}$ is defined to be $x_i$ and thus $(w^{-1})^{-1}=w.$  The \emph{empty word} (the word with no tokens) is denoted 1.

A word $w$ is  \emph{reduced} if, for all $i\in I$, the tokens $x_i$ and $x_i^{-1}$ never occupy adjacent positions in $w$.  E.g., the empty word is reduced, and if $w$ is reduced then so is $w^{-1}$.  Given any word $w$, one can ``simplify'' $w$ to a reduced word $w'$ by ``cancelling'' (i.e., deleting) all adjacent pairs of the form $x_i x_i^{-1}$ or  $ x_i^{-1} x_i$, then scanning the remaining tokens for other  such pairs in need of cancellation, etcetera, until a reduced word is obtained.  

The \emph{juxtaposition} of an ordered pair of reduced words,
\[  v = s_1s_2\ldots s_m\quad\mbox{and}\quad w = t_1t_2\ldots t_n, \]
is the concatenation of their token strings, denoted
\[  (v,w) \equiv s_1s_2\ldots s_m t_1t_2\ldots t_n. \]
If $t_1=s_m^{-1}$ then this new word, $(v,w)$, is not  a \emph{reduced} word, so define the  product of two reduced words to be the \emph{simplified} juxtaposition of their token strings,
\begin{equation}\label{reduced_word_product}
vw \equiv (v,w)'.
\end{equation}
The empty word, 1, is an identity element for (\ref{reduced_word_product}).  The technical crux in proving that (\ref{reduced_word_product})  defines a  group  is verifying  the associative law.  An additional argument then shows that the reduced-word group satisfies Definition~\ref{defn:FreeGroups}; see~\cite[Section~I.9]{Hungerford74}, \cite[Chapter~11]{Rotman95}, or~\cite[Chapter~2]{Robinson96} for the details.

\subsection{Free Products of Groups}\label{sec:Free:Products}
\begin{figure}[tb]
  \begin{center}
    \includegraphics{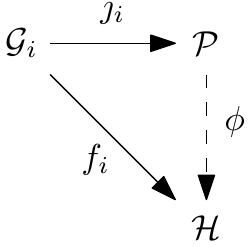}
    \caption{Commutative diagram defining a free product of the groups $\mathscr{G}_i$.}
    \label{fig:Coproduct}
  \end{center}
\end{figure}

Instead of seeking a  group that is freely generated by a given set of generators $x_i$, we now define a  group $\mathscr{P}$ that is ``freely'' generated by a given set of \emph{groups} $\mathscr{G}_i$.  
\begin{defn}[Free products~\cite{Hungerford74,Rotman95,Robinson96}]\label{defn:Coproducts}
Let $\{\mathscr{G}_i\colon i\in I\}$ be an indexed family of groups and let $\mathscr{P}$ be a group with homomorphisms $\jmath_i\colon\mathscr{G}_i\rightarrow \mathscr{P}$.  Then $\mathscr{P}$ is  a \emph{free product of the groups $\mathscr{G}_i$} if and only if, for every group $\mathscr{H}$ and family of homomorphisms $f_i\colon\mathscr{G}_i\rightarrow \mathscr{H}$, there exists a unique homomorphism
\[ \phi:\mathscr{P}\rightarrow \mathscr{H} \]
such that $\phi\circ \jmath_i=f_i$ for all $i\in I$.  This  is equivalent to saying that there exists a unique homomorphism $\phi$ such that the diagram in Figure~\ref{fig:Coproduct} commutes for all $i\in I$.
\end{defn}

 \subsubsection{Properties}\label{sec:Free:Products:Properties}
This is another ``categorical'' definition; the  category-theoretic name for an object $\mathscr{P}$ satisfying the  universal mapping property  in Figure~\ref{fig:Coproduct} is a \emph{coproduct}. 
For instance, in a category of vector spaces the weak direct sum $\sum V_i$ of a  set of vector spaces is a coproduct of the $V_i$.    The formal connection with free groups follows from Definitions~\ref{defn:FreeGroups} and~\ref{defn:Coproducts}.

\begin{prop}\label{prop:FreeGroupsAreCoproducts}
$\mathscr{F}$ is  free  on a set $I$ if and only if $\mathscr{F}$ is  a free product of infinite cyclic groups $\langle x_i \rangle$, indexed by  $i\in I$.
\end{prop}

The generators (or factors) $\mathscr{G}_i$ of a free product, $\mathscr{P}$, are groups with their own internal structure, and the homomorphisms $\jmath_i$ in a free product are injective~\cite[Lemma~11.49]{Rotman95}, so $\mathscr{P}$ contains isomorphic copies of   the factor groups $\mathscr{G}_i$.  As with free groups, abstract nonsense implies that free products are uniquely determined up to isomorphism by their generators~\cite[Theorem~I.7.5]{Hungerford74},   \cite[Theorem~11.50]{Rotman95}. 

\begin{prop}\label{prop:CoproductUniqueness}
Let $\{\mathscr{G}_i:i\in I\}$ be a collection of groups.  If $\mathscr{P}$ and $\mathscr{P}'$ are both free products of the groups $\mathscr{G}_i$ then $\mathscr{P}\cong\mathscr{P}'$.
\end{prop}

\subsubsection{Constructing free products}\label{sec:Free:Products:Construct}
The reduced word construction can be adapted, with a few modifications, to construct a canonical free product of an arbitrary collection of  groups, $\mathscr{G}_i$, $i\in I$.  The  token alphabet in the case of free products is defined to be the (disjoint) union of the factor groups: $X=\dotcup\mathscr{G}_i$.  The $\mathscr{G}_i$ are groups so $X$ is closed under inversion: $x\in X$ implies $x^{-1}\in X$.  There are also many more opportunities for simplification than just cancelling adjacent pairs of the form $x_i x_i^{-1}$.  A word  on $X$ is  \emph{reduced} if (1) two tokens from the same $\mathscr{G}_i$ never occupy adjacent positions, and (2) none of the tokens  is an identity element from any of the  $\mathscr{G}_i$.  Given any word $w$, one can simplify $w$ to a reduced word, $w'$, by multiplying all pairs of adjacent tokens  from the same group, deleting all  identity elements,  scanning  for other  tokens in need of simplification, etcetera, until a reduced word is obtained.    A  product  for reduced words is defined as the simplified juxtaposition of token strings.  Arguments similar to the ones for free groups prove that this product is associative and  that the resulting group of reduced words satisfies Definition~\ref{defn:Coproducts}; see~\cite[Theorem~I.9.6]{Hungerford74},   \cite[Theorem~11.51]{Rotman95},   \cite[Theorem~6.2.2]{Robinson96}.

A commonly used hieroglyph for free products is $\bigast$,  e.g.,
\[ \mathscr{P} = \mathscr{G}_1\bigast\mathscr{G}_2. \]
We will use this notation specifically to denote the free product realization given by the reduced word construction.  
Definition~\ref{defn:Coproducts} and the reduced word construction are insensitive to the order in which the  groups  $\mathscr{G}_i$ are indexed, so the operator $\bigast$ is trivially commutative; i.e.,
\[  \mathscr{G}_1\bigast\mathscr{G}_2 = \mathscr{G}_2\bigast\mathscr{G}_1. \]
This should {not} be confused with the fact (which follows from the reduced word construction) that $\mathscr{G}_1\bigast\mathscr{G}_2$ is a  \emph{nonabelian} group.   For instance, if $g_1\in\mathscr{G}_1$ and $g_2\in\mathscr{G}_2$ then  $g_1 g_2$ and $g_2 g_1$ are \emph{different} reduced words:
\[  g_1 g_2 \neq g_2 g_1.  \]
This is true even when the individual factor groups $\mathscr{G}_i$ are abelian, which is the case of interest in this paper.

\subsubsection{Connection with lifting cascade groups}\label{sec:Free:Products:Lifting}
What does the reduced word construction of free products have to do with  lifting cascade groups?    Given a group lifting structure \mbox{$(\mathscr{D},\mathscr{U},\mathscr{L},\mathfrak{B})$} with upper and lower triangular lifting matrix groups $\mathscr{U}$ and $\mathscr{L}$, the lifting cascade group $\mathscr{C}$ is the  group  generated by $\mathscr{U}$ and $\mathscr{L}$. Although the string  $\mathbf{S}_N\cdots\mathbf{S}_1$ in~(\ref{lifting_cascade_group}) represents the \emph{product}  of the lifting matrices $\mathbf{S}_i(z)$, the lifting cascade group  is clearly in a one-to-many correspondence with the set of all \emph{words}  on the alphabet $\mathscr{U\cup L}$.  To eliminate  degenerate, trivially nonunique lifting factorizations, the author created an \emph{ad hoc} definition of ``irreducible'' lifting cascades~\cite[Definition~3]{Bris:10:GLS-I} (see Section~\ref{sec:Intro:Notation:Polyphase} above).

The question of whether transfer  matrices in $\mathscr{C}$ and \emph{reduced} words on the alphabet $\mathscr{U\cup L}$ are in one-to-one correspondence sounds a lot like asking whether matrices in $\mathscr{C}$ have unique \emph{irreducible} lifting factorizations over $\mathscr{U}$ and $\mathscr{L}$.  This in turn is very close to the ``uniqueness-modulo-rescaling'' results  established in~\cite{Bris:10b:GLS-II} for the two nontrivial classes of linear phase filter banks.  The pain inflicted by reading \cite{Bris:10b:GLS-II} indicates just how far irreducibility is from being  \emph{sufficient}  for  uniqueness of lifting factorizations.

While the ``correspondence'' just described between a lifting cascade group, $\mathscr{C}$, and a reduced word realization of a free product, $\mathscr{U\bigast L}$,  
is highly suggestive, the subject matter is sufficiently technical that a formal proof is needed to show that $\mathscr{C}$ is a free product of $\mathscr{U}$ and $\mathscr{L}$.  We  prove directly that lifting cascade groups with unique irreducible group lifting factorizations satisfy Definition~\ref{defn:Coproducts} without assuming any  results on the existence of canonical free products.  Modulo  the technicalities behind the reduced word construction it then follows from  Proposition~\ref{prop:CoproductUniqueness} that $\mathscr{C}\cong\mathscr{U\bigast L}$.

\subsection{Structure of Lifting Cascade Groups}\label{sec:Free:Cascade}
\subsubsection{Free product structure}\label{sec:Free:Cascade:Product}
\begin{figure}[tb]
  \begin{center}
    \includegraphics{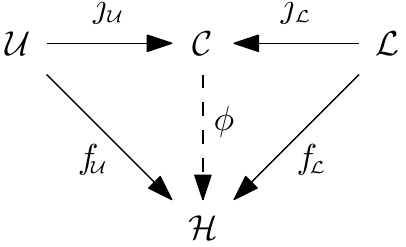}
    \caption{Universal mapping property for the coproduct $\mathscr{C\cong U \bigast L}$.}
    \label{fig:UstarL}
  \end{center}
\end{figure}

Our unique factorization tool~\cite[Theorem~1]{Bris:10:GLS-I} is based on  group lifting structures $(\mathscr{D,U,L},\mathfrak B)$.  Since a  lifting cascade group, $\mathscr{C\equiv\langle U\cup L\rangle}$, does not depend on $\mathscr D$ or $\mathfrak B$, neither does the statement of Theorem~\ref{thm:Cascade} (below).  The phenomenon of uniqueness modulo rescaling in the conclusion of~\cite[Theorem~1]{Bris:10:GLS-I} is  addressed by the next lemma, which implies \emph{uniqueness} of irreducible group lifting factorizations in $\mathscr C$  whenever~\cite[Theorem~1]{Bris:10:GLS-I} holds, even if factorizations in $\mathscr{DC}\mathfrak B$ are only unique modulo rescaling.
\begin{lem}\label{lem:Cascade}
If $\mathfrak{S}$ is a $\mathscr{D}$-invariant, order-increasing group lifting structure with lifting cascade group $\mathscr{C\equiv\langle U\cup L\rangle}$ then irreducible group lifting factorizations in $\mathscr C$ are unique.
\end{lem}
\pf
Suppose we are given two irreducible group lifting factorizations of $\mathbf{E}(z)\in\mathscr{C}$ with $\mathbf{S}_i(z),\,\mathbf{S}'_i(z)\in\mathscr{U\cup L}$:
\begin{eqnarray}
\mathbf{E}(z) & = & \mathbf{S}_{N-1}(z) \cdots\mathbf{S}_0(z)   \label{unprimed_cascade} \\
        & = & \mathbf{S}'_{N'-1}(z)\cdots \mathbf{S}'_0(z)\,. \label{primed_cascade}
\end{eqnarray}
The base matrices  are \mbox{$\mathbf{B}(z)=\mathbf{I}=\mathbf{B}'(z)$} so $\mathbf{E}(z)\in\mathscr{DC}\mathfrak B$ if and only if $\mathbf{I}\in\mathfrak B$.  If $\mathbf{I}\notin\mathfrak B$ define a new matrix $\mathbf{G}(z)\in\mathscr{DC}\mathfrak B$ using any $\mathbf{B}(z)\in\mathfrak B$ and cascades (\ref{unprimed_cascade}) and (\ref{primed_cascade}):
\begin{eqnarray}
\mathbf{G}(z) & \equiv & \mathbf{S}_{N-1}(z) \cdots\mathbf{S}_0(z)\, \mathbf{B}(z)    \label{unprimed_G_cascade} \\
        & = & \mathbf{S}'_{N'-1}(z)\cdots \mathbf{S}'_0(z)\, \mathbf{B}(z) \,. \label{primed_G_cascade}
\end{eqnarray}
In either case, the scaling matrices  are \mbox{$\mathbf{D}_K=\mathbf{I}=\mathbf{D}_{K'}$} so application of~\cite[Theorem~1]{Bris:10:GLS-I} to 
(\ref{unprimed_cascade})--(\ref{primed_cascade}) or 
(\ref{unprimed_G_cascade})--(\ref{primed_G_cascade}) 
shows that $N'=N$ and $\mathbf{S}'_i(z)  =  \mathbf{S}_i(z)$ for $i=0,\ldots,N-1$. \hfill\qed
\begin{thm}[Lifting cascade group structure]
\label{thm:Cascade}
Let $\mathscr{U}$ and $\mathscr{L}$ be  lifting matrix groups with lifting cascade group  $\mathscr{C\equiv\langle U\cup L\rangle}$.
If every element of $\mathscr{C}$ has a unique irreducible group lifting factorization over $\mathscr{U\cup L}$ then $\mathscr{C}$ is a free product of $\mathscr{U}$ and $\mathscr{L}$ and is therefore isomorphic to the reduced word realization,  
\[ \mathscr{C\cong U \bigast L}. \]
\end{thm}
\pf
We  show that $\mathscr C$ satisfies Definition~\ref{defn:Coproducts}. Let $\jmath_{\ssst\mathscr U}$ and  $\jmath_{\ssst\mathscr L}$ be the inclusion isomorphisms of $\mathscr U$ and $\mathscr L$ into $\mathscr C$.  Suppose we are given a group $\mathscr H$ and homomorphisms 
\[  f_{\!\ssst\mathscr U}\colon \mathscr{U}\rightarrow \mathscr{H}\quad\mbox{and}\quad 
f_{\!\ssst\mathscr L} \colon \mathscr{L}\rightarrow \mathscr{H}.  \]
We need to show that there exists a unique homomorphism, 
\[  \phi\colon \mathscr{C}\rightarrow \mathscr{H},  \]
such that the diagram in Figure~\ref{fig:UstarL} commutes.

From now on, identify  $\mathscr U$ and $\mathscr L$ with their isomorphic images in  $\mathscr C$ under the inclusions $\jmath_{\ssst\mathscr U}$ and  $\jmath_{\ssst\mathscr L}$.  To make $\phi$ agree with $f_{\!\ssst\mathscr U}$ and $f_{\!\ssst\mathscr L}$ on  $\mathscr{U,\, L< C}$, define
\begin{equation}\label{phi_on_UuL}
\phi(\mathbf{S}) \equiv \left\{
\begin{array}{ll}
f_{\!\ssst\mathscr U}(\mathbf{S}),  & \mathbf{S}\in\mathscr{U}, \\
f_{\!\ssst\mathscr L}(\mathbf{S}),  & \mathbf{S}\in\mathscr{L}.
\end{array}
\right.
\end{equation}
$\mathscr{U\cap L}=\mathbf{I}$ and  $f_{\!\ssst\mathscr U}(\mathbf{I})=1_{\mathscr H}=f_{\!\ssst\mathscr L}(\mathbf{I})$  so (\ref{phi_on_UuL}) is well-defined.
Extend $\phi$ to a function on all of $\mathscr{C}$: if $\mathbf{E}=\mathbf{S}_N\cdots\mathbf{S}_0$ is the unique irreducible group lifting factorization of $\mathbf{E}\in\mathscr{C}$, define
\begin{equation}\label{phi_on_C}
\phi(\mathbf{E}) \equiv \phi(\mathbf{S}_N)\cdots\phi(\mathbf{S}_0),
\end{equation}
where $\phi(\mathbf{S}_i)$ is given by (\ref{phi_on_UuL}).  The associative law in $\mathscr{H}$ and uniqueness of irreducible group lifting factorizations  imply that (\ref{phi_on_C}) is well-defined; we must show  it is a homomorphism.

Let   $\mathbf{E},\,\mathbf{E'}\in\mathscr{C}$ and let $N$ (respectively, $N'$) be the lengths of their unique irreducible group lifting factorizations,
\[ \mathbf{E}=\mathbf{S}_{N-1}\cdots\mathbf{S}_0\quad\mbox{and}\quad
\mathbf{E'}=\mathbf{S}'_{N'-1}\cdots\mathbf{S}'_0.  \]
We will prove that
\begin{equation}\label{homomorphism}
\phi(\mathbf{E}) \phi(\mathbf{E'}) = \phi(\mathbf{EE'})
\end{equation}
by induction on $N_{\rm tot}\equiv N+N'$.  Property~(\ref{homomorphism}) is trivial if either matrix is $\mathbf I$, so we always assume that $N,\,N'\geq 1.$

\case{$N_{\rm tot}=2$ ($N,\,N'=1$).} We are given $\mathbf{E}=\mathbf{S}_0$ and $\mathbf{E}'=\mathbf{S}'_0$.  If $\mathbf{S}_0$ and $\mathbf{S}'_0$ have opposite update characteristics then $\mathbf{S}_0\mathbf{S}'_0$ is the (unique) irreducible group lifting factorization of $\mathbf{E}\mathbf{E}'$ and~(\ref{homomorphism}) is just definition~(\ref{phi_on_C}) for $\phi(\mathbf{EE'})$.  If $\mathbf{S}_0$ and $\mathbf{S}'_0$ have the \emph{same} update characteristic (i.e., $\mathbf{S}_0,\,\mathbf{S}'_0\in\mathscr{G}$ for $\mathscr{G}=\mathscr{U}$ or $\mathscr{G}=\mathscr{L}$), then 
$\phi(\mathbf{S}_0)=f_{\!\ssst\mathscr{G}}(\mathbf{S}_0)$ and $\phi(\mathbf{S}'_0)=f_{\!\ssst\mathscr{G}}(\mathbf{S}'_0)$  so
(\ref{homomorphism}) is the homomorphism property of $f_{\!\ssst\mathscr{G}}$.

\case{$N_{\rm tot}>2.$}  Assume (\ref{homomorphism}) holds for  products in which $N+N'<N_{\rm tot}$, and let $\mathbf{E},\,\mathbf{E'}\in\mathscr{C}$ have  irreducible group lifting factorizations with  $N+N'=N_{\rm tot}$ lifting steps.   

If $\mathbf{S}_0$ and $\mathbf{S}'_{N'-1}$ have opposite update characteristics then 
\[  \mathbf{EE'}=\mathbf{S}_{N-1}\cdots\mathbf{S}_1\mathbf{S}_0\mathbf{S}'_{N'-1}\mathbf{S}'_{N'-2}\cdots\mathbf{S}'_0 \]
is the irreducible group lifting factorization of $\mathbf{EE'}$ so, by associativity in $\mathscr{H}$,
\begin{eqnarray*}
\phi(\mathbf{E}) \phi(\mathbf{E'}) 
	& = & (\phi(\mathbf{S}_{N-1})\cdots\phi(\mathbf{S}_0))\cdot(\phi(\mathbf{S}'_{N'-1})\cdots\phi(\mathbf{S}'_0)) \\
	& = & \phi(\mathbf{S}_{N-1})\cdots\phi(\mathbf{S}_0)\phi(\mathbf{S}'_{N'-1})\cdots\phi(\mathbf{S}'_0)  \\
	& = & \phi(\mathbf{EE'})\quad\mbox{by~(\ref{phi_on_C}).}
\end{eqnarray*}

If $\mathbf{S}_0$ and $\mathbf{S}'_{N'-1}$ have the \emph{same} update characteristic, i.e., $\mathbf{S}_0,\,\mathbf{S}'_{N'-1}\in\mathscr{G}$ for $\mathscr{G}=\mathscr{U}$ or $\mathscr{G}=\mathscr{L}$, let $\mathbf{S}' \equiv\mathbf{S}_0\mathbf{S}'_{N'-1}\in\mathscr{G}$.  By associativity in $\mathscr{H}$ and the homomorphism property of $f_{\!\ssst\mathscr{G}}$,
\begin{eqnarray}
\phi(\mathbf{E}) \phi(\mathbf{E'}) 
	& = & \phi(\mathbf{S}_{N-1})\cdots\phi(\mathbf{S}_0)\phi(\mathbf{S}'_{N'-1})\cdots\phi(\mathbf{S}'_0) \nonumber \\
	& = & \phi(\mathbf{S}_{N-1})\cdots\phi(\mathbf{S}')\cdots\phi(\mathbf{S}'_0). \label{homo_step1}
\end{eqnarray}
Write $\mathbf{EE}'=\mathbf{VW}$ for the  irreducible group lifting cascades
\begin{equation}\label{reduced_factorization}
\mathbf{V}\equiv\mathbf{S}_{N-1}\cdots\mathbf{S}_1\quad\mbox{and}\quad
\mathbf{W}\equiv\mathbf{S}' \mathbf{S}'_{N'-2} \cdots\mathbf{S}'_0.
\end{equation}
(Note that $\mathbf{V}=\mathbf{I}$ if $N=1$.  Similarly, $\mathbf{W}=\mathbf{I}$ if $\mathbf{S}' = \mathbf{I}$ and $N'=1$.)  Reassociate factors in~(\ref{homo_step1}) and use irreducibility of the cascades in~(\ref{reduced_factorization}) to get
\begin{equation}\label{homo_step2}
\phi(\mathbf{E}) \phi(\mathbf{E'}) = \phi(\mathbf{V}) \phi(\mathbf{W}) .
\end{equation}
If either $\mathbf{V}=\mathbf{I}$  or $\mathbf{W}=\mathbf{I}$ then the right-hand side of (\ref{homo_step2}) trivially reduces to 
\begin{equation}\label{homo_step3}
\phi(\mathbf{V}) \phi(\mathbf{W}) = \phi(\mathbf{VW}).  
\end{equation}
If neither $\mathbf{V}$  nor $\mathbf{W}$ is $\mathbf{I}$ then the total number of lifting matrices in~(\ref{reduced_factorization}) for $\mathbf{V}$ and $\mathbf{W}$ is at most $N_{\rm tot}-1$, and applying the induction hypothesis to $\mathbf{V}$ and $\mathbf{W}$ yields (\ref{homo_step3}). In any case,
\begin{equation}\label{homo_step4}
\phi(\mathbf{E}) \phi(\mathbf{E'}) = \phi(\mathbf{VW}) = \phi(\mathbf{EE}') .
\end{equation}

This proves that $\phi$ is a homomorphism. Uniqueness of $\phi$ is straightforward: since $\mathscr{U\cup L}$ generates $\mathscr{C}$, definitions~(\ref{phi_on_UuL}) and~(\ref{phi_on_C}) show that any  homomorphism $\psi:\mathscr{C}\rightarrow\mathscr{H}$ extending $f_{\!\ssst\mathscr U}$ and $f_{\!\ssst\mathscr L}$ necessarily agrees with $\phi$ on all of $\mathscr{C}$.  We have therefore shown that $\mathscr{C}$ is a free product of $\mathscr{U}$ and $\mathscr{L}$.

According to ~\cite[Theorem~I.9.6]{Hungerford74},   \cite[Theorem~11.51]{Rotman95},   \cite[Theorem~6.2.2]{Robinson96} the reduced word construction also yields a free product, which we have been denoting $\mathscr{U \bigast L}$, so by Proposition~\ref{prop:CoproductUniqueness} this implies that $\mathscr C$ is isomorphic to  $\mathscr{U \bigast L}$.
\hfill\qed

\rem  By \cite[Theorem~1 and Theorem~2]{Bris:10b:GLS-II} and Lemma~\ref{lem:Cascade},  the lifting cascade groups for the   WS and HS group lifting structures  have unique irreducible group lifting factorizations, so  Theorem~\ref{thm:Cascade} implies that they are free products of their lifting matrix groups.

The converse of Theorem~\ref{thm:Cascade} is also true, meaning that the free product representation $\mathscr{C\cong U \bigast L}$ is \emph{equivalent} to uniqueness of irreducible group lifting factorizations. The proof follows from uniqueness of reduced word representations in canonical  free products~\cite[Theorem~11.52]{Rotman95}, \cite[Theorem~6.2.3]{Robinson96}.

\begin{prop}\label{prop:CascadeEquivalence}
Let $\mathscr{C\equiv\langle U\cup L\rangle}$;  if $\mathscr{C\cong U \bigast L}$ then irreducible group lifting factorizations in $\mathscr{C}$ are unique.
\end{prop}

\subsubsection{Free lifting cascade groups}\label{sec:Free:Cascade:Free}
In light of Theorem~\ref{thm:Cascade} it is natural to ask whether we can characterize the group lifting structures for which $\mathscr{C\equiv\langle U\cup L\rangle}$ is  a \emph{free group} and not just a free \emph{product.}
In one direction, we will show (Theorem~\ref{thm:FreeCascadeGroup}) that  if $\mathscr U$ and $\mathscr L$ are infinite cyclic groups and $\mathscr{C\cong U \bigast L}$  (which are highly restrictive hypotheses) then it follows easily that $\mathscr{C}$ is  free, but this implication does \emph{not} hold without the condition $\mathscr{C\cong U \bigast L}$.  The converse  (i.e., the \emph{necessity} of having a free product of two infinite cyclic subgroups)  follows from some basic facts about  $\mathscr U$ and $\mathscr L$ and some nontrivial group theory.  

The need to combine  infinite cyclic  groups using something as complicated as a free product in order to get a free group  can  be understood in light of the theory of \emph{group presentations}~\cite[Section~I.9]{Hungerford74},   \cite[Chapter~11]{Rotman95}, \cite[Chapter~2]{Robinson96}, in which a free group is  distinguished by having a set of ``free'' generators---ones that do not satisfy any  \emph{relations} (factorizations of the identity).   In previous work~\cite[Equation~(4) and Example~1]{Bris:10:GLS-I}  we  presented irreducible liftings of the identity  as  obstructions to uniqueness of lifting factorizations, and we now show that such relations can arise from as few as two  generator matrices.

\begin{exmp}\label{exmp:NonFreeCascadeGroup}
Let  $\mathscr U$, $\mathscr L$  be infinite cyclic groups,  $\mathscr{U}=\langle\mathbf{S}_0(z)\rangle$ and $\mathscr{L}=\langle\mathbf{S}_1(z)\rangle$,  generated by  lifting filters $S_0(z)$ and $S_1(z)$:
\[  \mathbf{S}_0(z)\equiv\upsilon(S_0(z)),\quad \mathbf{S}_1(z)\equiv\lambda(S_1(z)).  \]
Let $\mathscr{C\equiv\langle U\cup L\rangle}$ be the associated  lifting cascade group.  In spite of having noncommuting generator matrices $\mathbf{S}_0(z)$ and $\mathbf{S}_1(z)$,
$\mathscr{C}$ may nonetheless fail to be a \emph{free} group because of a  relation involving  $\mathbf{S}_0(z)$ and $\mathbf{S}_1(z)$.  E.g., consider
\[  S_0(z) \equiv az^{-d}\quad\mbox{and}\quad S_1(z) \equiv -a^{-1}z^d;\quad a\neq 0,\; d\in\mathbb{Z} . \] 
The reader can verify that the corresponding lifting matrices  satisfy the inobvious relation
\begin{equation}\label{torsion}
(\mathbf{S}_0(z)\mathbf{S}_1(z))^6 = \mathbf{I}.  
\end{equation}
This shows that  $\mathbf{S}_0(z)$ and $\mathbf{S}_1(z)$ are not \emph{free} generators for $\mathscr{C}$.  Moreover,  using the notion of \emph{cyclically reduced words}~\cite[p.~434]{Rotman95} and uniqueness of spelling for reduced words, one can show that free groups never contain elements of finite, nonzero order.  Since~(\ref{torsion}) says that the product $\mathbf{S}_0(z)\mathbf{S}_1(z)$ has order~6, $\mathscr{C}$ cannot be a free group on \emph{any} set of generators.

Note that the WS and HS group lifting structures  do \emph{not} include monomial lifting filters, although~\cite[Example~1]{Bris:10:GLS-I} shows that excluding monomial lifting filters is far from sufficient for ensuring unique irreducible group lifting factorizations.
\end{exmp}

\begin{thm}\label{thm:FreeCascadeGroup}
Let $\mathscr{C\equiv\langle U\cup L\rangle}$ be a lifting cascade group  over nontrivial lifting matrix groups  $\mathscr U$ and $\mathscr L$.   $\mathscr{C}$ is a free group  if and only if $\mathscr U$ and $\mathscr L$  are infinite cyclic  and $\mathscr{C\cong U \bigast L}$.
\end{thm}
\pf
If $\mathscr U$ and $\mathscr L$  are infinite cyclic groups and $\mathscr{C\cong U \bigast L}$ then $\mathscr{C}$ is free on two generators by  Proposition~\ref{prop:FreeGroupsAreCoproducts}.

Conversely, suppose  $\mathscr{C\equiv\langle U\cup L\rangle}$ is a free group.  Since $\mathscr{U}$ and $\mathscr{L}$ are subgroups of $\mathscr{C}$, both $\mathscr{U}$ and $\mathscr{L}$ are free  by the Nielsen-Schreier Theorem~\cite[Theorem~7.2.1]{Hall99}, \cite[Theorem~11.44]{Rotman95}.  Lifting matrix groups are  abelian, but the free groups with two or more generators constructed by the reduced word construction are  \emph{nonabelian} so Proposition~\ref{prop:FreeGroupUniqueness}  implies that $\mathscr{U}$ and $\mathscr{L}$ cannot be free on two or more generators.  Therefore, they must be free on just one generator apiece; i.e., infinite cyclic groups.  Since $\mathscr{C}$ is generated by $\mathscr{U}$ and $\mathscr{L}$ it is free on two generators, so Proposition~\ref{prop:FreeGroupsAreCoproducts} implies  that $\mathscr{C\cong U \bigast L}$.
\hfill\qed


\section{Semidirect Product Structure of the Scaled Lifting Group}\label{sec:Semidirect}
Our next topic is the group-theoretic structure of the scaled lifting group, $\mathscr S$.  The relationship between the gain-scaling group, $\mathscr D$,  and the lifting cascade group, $\mathscr C$, both of which are subgroups of $\mathscr S$ by definition~(\ref{scaled_lifting_group}), is characterized by a construction called a \emph{semidirect product}~\cite{MacLaneBirkhoff67,Rotman95,Robinson96,Hall99}.

\subsection{Semidirect Products of Groups}\label{sec:Semidirect:Groups}

\begin{defn}[Semidirect products]\label{defn:SemidirectProduct}
Let $\mathscr G$ be a (multiplicative) group with identity element $1_{\mathscr G}$ and subgroups $\mathscr K$ and $\mathscr Q$.  
$\mathscr G$ is the \emph{(internal) semidirect product of $\mathscr K$ by $\mathscr Q$}, denoted $\mathscr{G=Q\ltimes K}$, if the following three axioms are satisfied. 
\begin{eqnarray}
&&\mathscr{G=\langle K\cup Q\rangle}\quad \mbox{($\mathscr K$ and $\mathscr Q$ generate $\mathscr G$)}\label{G=KvQ}\\
&&\mathscr{K\lhd G}\quad  \mbox{($\mathscr K$ is a normal subgroup of $\mathscr G$)}\label{normal_subgp}\\
&&\mathscr{K\cap Q} = 1_{\mathscr G}\quad\mbox{(the trivial group)}\label{trivial_gp}
\end{eqnarray}
\end{defn}

\subsubsection{Product Representations}\label{sec:Semidirect:Groups:Product}
Let  $\mathscr G$ be  generated by subgroups $\mathscr{K}$ and  $\mathscr{Q}$.  If $\mathscr{K\lhd G}$ then $\mathscr{Q}$ acts on $\mathscr{K}$ by inner automorphisms  (``$\mathscr{K}$ is $\mathscr{Q}$-invariant'' \cite{Bris:10:GLS-I},  ``$\mathscr{Q}$ normalizes $\mathscr{K}$''~\cite{Rotman95}):
\begin{equation}\label{Q_action}
\gamma\colon {\mathscr Q}\rightarrow\Aut({\mathscr K})\quad\mbox{where}\quad 
\gamma_q k \equiv qkq^{-1}\in\mathscr{K}.
\end{equation}
(Note that we have extended  definition (\ref{defn_gamma}) for $\gamma$  to the abstract group-theoretic setting.)   The converse also holds.

\begin{lem}\label{lem:Q_action}
If $\mathscr{G=\langle K\cup Q\rangle}$ and  $\mathscr{Q}$ acts on $\mathscr{K}$ via inner automorphisms (\ref{Q_action})  then $\mathscr{G}= \mathscr{QK}$ and $\mathscr{K\lhd G}$.  If~(\ref{trivial_gp}) also holds (i.e., if $\mathscr{G=Q\ltimes K}$) then the product representations in $\mathscr{G}= \mathscr{QK}$ are unique.
\end{lem}
\pf
Any $g\in\mathscr{G}=\langle\mathscr{K\cup Q}\rangle$ can be written 
\begin{equation}\label{prod_g_i}
g=g_0\cdots g_n\,,\quad g_i\in\mathscr{K\cup Q}\,. 
\end{equation}
Products of the form $g_i g_{i+1}=kq$ can be written $kq = qk'$ using (\ref{Q_action}),
 so~(\ref{prod_g_i}) can be rewritten $g=(q_0q_1\cdots)(k_0'k_1'\cdots)$, implying
\begin{equation}\label{G=QK}
\mathscr{G=QK}.
\end{equation}
Normality of $\mathscr{K}$ in $\mathscr{G}$ follows easily using (\ref{Q_action}) and (\ref{G=QK}), and (\ref{trivial_gp}) implies uniqueness in (\ref{G=QK})  since $q_0k_0  = q_1k_1$  implies 
\begin{equation}\label{unique_prod_rep}
k_0k_1^{-1}  =  q_0^{-1}q_1 \in\mathscr{K\cap Q} = 1_{\mathscr G}.
\end{equation}
\qed

\rem  We use the notation $Q\ltimes K$ for semidirect products (compared to the more common $K\rtimes Q$, e.g.~\cite{Rotman95}) because the  convention in lifting is to put the  scaling matrix $\mathbf{D}_K$ at the \emph{left} end of the analysis cascade~(\ref{lift_from_B}).  This corresponds to the product representation $\mathscr{S=DC}$ for $\mathscr{D}$-invariant group lifting structures~\cite[formula~(31)]{Bris:10:GLS-I}, which is of the form~(\ref{G=QK}).

\subsubsection{External semidirect products}\label{sec:Semidirect:Groups:External}
If $\mathscr{G=Q\ltimes K}$ then a ``twisted multiplication'' formula holds for  $\mathscr{G=QK}$ (cf.~\cite{Rotman95}):
\begin{equation}\label{twisted_mult}
\begin{array}{rcccl}
g_0g_1 & = & (q_0k_0)(q_1k_1) & = & q_0q_1(\gamma_{\!q_1^{-1}}k_0)k_1,\\
g^{-1} & = & (qk)^{-1} & = & q^{-1}\gamma_{\!q}k^{-1} .
\end{array}
\end{equation}

Note how the formulas in (\ref{twisted_mult}) represent $g_0g_1$ and $g^{-1}$ as factored elements of $\mathscr{QK}$.  This leads to an alternate definition of semidirect product that does not require $\mathscr K$ and $\mathscr Q$ to be subgroups of a common parent.  Given any homomorphism 
\[  \theta\colon {\mathscr Q}\rightarrow\Aut({\mathscr K}),  \]
one mimics~(\ref{twisted_mult}) using  $\theta$ in place of $\gamma$ to \emph{define} an associative twisted multiplication on the cartesian product $\mathscr{Q\times K}$ called the \emph{(external) semidirect product of $\mathscr K$ by $\mathscr Q$}, denoted $\mathscr{Q\ltimes_{\theta} K}$.  This is \emph{not} the same as the more familiar ``direct'' product of  $\mathscr Q$ and $\mathscr K$ given by the set $\mathscr{Q\times K}$ with coordinate-wise multiplication.  The external semidirect product thus defined on the set $\mathscr{Q\times K}$ makes $\mathscr{Q\ltimes_{\theta} K}$ the \emph{internal} semidirect product of $1_{\mathscr Q}\times {\mathscr K}$ by ${\mathscr Q}\times 1_{\mathscr K}$.  In this twisted product, the twisted conjugation of $(1_{\mathscr Q},k)$ by $(q,1_{\mathscr K})$ is given by the automorphism $1\times\theta_q\in \Aut(1_{\mathscr Q}\times{\mathscr K})$; i.e., 
\[  (q,1_{\mathscr K})(1_{\mathscr Q},k)(q,1_{\mathscr K})^{-1} = (1_{\mathscr Q},\theta_q k)\quad\mbox{for all $k\in\mathscr{K}$.} \]

For example, wreath products 
\cite{FooteMirchandEtal:00:Wreath-Product-Group,MirchandFooteEtal:00:Wreath-Product-Group,FooteMirchandEtal:04:Two-Dimensional-Wreath-Product} 
are  defined in terms of external semidirect products.  The product formula  in
\cite[Section~III-B.1]{FooteMirchandEtal:00:Wreath-Product-Group} 
can be interpreted as a twisted multiplication for a certain automorphic group action~\cite[Chapter~7]{Rotman95}.

Suppose that $\mathscr{G=Q\ltimes K}$ and we are given an isomorphism
\[  \rho\colon\mathscr{K}\stackrel{\cong}{\longrightarrow}\mathscr{J}  \]
of $\mathscr{K}$ onto some group $\mathscr{J}$.  We want to translate $\mathscr{Q\ltimes K}$ into an equivalent \emph{external} semidirect product of $\mathscr{J}$ by $\mathscr{Q}$.  Use $\rho$ to push the automorphisms $\gamma_{\!q}$ from $\mathscr{K}$ onto $\mathscr{J}$ by  defining
\begin{equation}\label{def_theta}
\theta_{\!q}\equiv\rho\circ\gamma_{\!q}\circ\rho^{-1}\quad\mbox{for all $q\in\mathscr{Q}$.}  
\end{equation}
Since $\gamma_{\!q}$ is an automorphism of $\mathscr{K}$ and $\rho$ is an isomorphism, the composition $ \theta_{\!q}$ is an automorphism of $\mathscr{J}$.    Moreover, $\gamma$ is a homomorphism of $\mathscr{Q}$ into $\Aut(\mathscr{K})$ so we get a homomorphism
\begin{equation}\label{theta_automorphic}
\theta\colon\mathscr{Q}\rightarrow\Aut(\mathscr{J}), 
\end{equation}
which defines an external semidirect product $\mathscr{Q\ltimes_{\theta} J}$.

\begin{lem}\label{lem:ext_semidirect}
Let $\rho$ and $\theta$ be given as above.  Define
\[  \psi\colon \mathscr{G=Q\ltimes K=QK} \rightarrow \mathscr{Q\ltimes_{\theta} J}, \]
\begin{equation}  
\psi(qk) \equiv (q,\rho k).  \label{def_psi}
\end{equation}
Then $\psi$ is an isomorphism of $\mathscr{G=Q\ltimes K}$ onto $\mathscr{Q\ltimes_{\theta} J}$.
\end{lem}
\pf
Note that $\psi$ is well-defined since product representations in $\mathscr{G}= \mathscr{QK}$ are unique by Lemma~\ref{lem:Q_action}.
First show that $\psi$ is a homomorphism.
\begin{eqnarray*}
\psi((q_0k_0)(q_1k_1)) & = & \psi(q_0q_1(\gamma_{\!q_1^{-1}}k_0)k_1)\mbox{\ by (\ref{twisted_mult})}\\
	&=& \left(q_0q_1,\, \rho((\gamma_{\!q_1^{-1}}k_0)k_1)\right) \mbox{\ by (\ref{def_psi})}\\
	&=& \left(q_0q_1,\, \rho(\gamma_{\!q_1^{-1}}k_0)\rho k_1\right)\\
	&=& \left(q_0q_1,\, (\theta_{\!q_1^{-1}}j_0) j_1\right),\mbox{\,where $j_i\equiv\rho k_i$}\\
	&=& (q_0,j_0)(q_1,j_1),\mbox{\,the product in $\mathscr{Q\ltimes_{\theta} J}$}\\
	&=& \psi(q_0k_0)\,\psi(q_1k_1).
\end{eqnarray*}

Next, show that $\psi\colon\mathscr{G=QK} \rightarrow \mathscr{Q\ltimes_{\theta} J}$ is injective. The identity element in $\mathscr{Q\ltimes_{\theta} J}$ is $(1_{\mathscr Q},\,1_{\mathscr J})$; suppose that
\[  (1_{\mathscr Q},\,1_{\mathscr J}) = \psi(qk) = (q,\rho k).  \]
Then $q = 1_{\mathscr Q} = 1_{\mathscr G}$, while $\rho k = 1_{\mathscr J}$ implies $k = 1_{\mathscr K} = 1_{\mathscr G}$ since $\rho$ is injective. This means that $qk = 1_{\mathscr G}$, proving that $\psi$ is injective.

Finally, for any $(q,j)\in \mathscr{Q\ltimes_{\theta} J}$ let $k\equiv \rho^{-1}j\in\mathscr{K}$, which is well-defined since $\rho$ is surjective. Thus, $\psi(qk) =  (q,j)$, proving that $\psi$ is surjective.
\qed

\subsection{Structure of Scaled Lifting Groups}\label{sec:Semidirect:Scaled}
Let $\mathfrak S$ be a  group lifting structure with  scaled lifting group~$\mathscr{S}\equiv\langle\mathscr{C\cup D}\rangle$.  We now give sufficient conditions for $\mathscr S$ to be a semidirect product.  Although the hypotheses of Theorem~\ref{thm:Semidirect} are the same as those of~\cite[Theorem~1]{Bris:10:GLS-I}, we do not invoke  unique factorization  but, rather, prove Theorem~\ref{thm:Semidirect} directly from the hypotheses by verifying Definition~\ref{defn:SemidirectProduct}. 
\begin{thm}
\label{thm:Semidirect}
If $\mathfrak S$ is a $\mathscr D$-invariant, order-increasing group lifting structure then   $\mathscr S$ is the  semidirect product of $\mathscr C$ by $\mathscr D$,
\[ \mathscr{S=D\ltimes C}. \]
\end{thm}
\pf
Axiom~(\ref{G=KvQ}), $\mathscr{S}=\langle\mathscr{C\cup D}\rangle$, is true by definition.

Axiom~(\ref{normal_subgp}), $\mathscr{C\lhd S}$, follows from Lemma~\ref{lem:Q_action} by $\mathscr D$-invariance.

To prove Axiom~(\ref{trivial_gp}), $\mathscr{C\cap D}=\mathbf{I}$,  let   $\mathbf{E}\neq \mathbf{I}$ be any nontrivial element of $\mathscr C$.  Let $\mathbf{D}_K\in\mathscr{D}$  and   $\mathbf{B}\in\mathfrak{B}$.  We have 
\[ \order(\mathbf{D}_K\mathbf{B}) = \order(\mathbf{B}), \]
but the order-increasing property of $\mathfrak S$ implies
\[ \order(\mathbf{EB}) > \order(\mathbf{B})\,. \]
It follows that $\mathbf{E}\neq \mathbf{D}_K$, which proves that $\mathscr{C\cap D}=\mathbf{I}$. \qed

\rem 
Axiom~(\ref{trivial_gp}) implies that 
$\mathscr U$ and $\mathscr L$ do \emph{not} generate lifting factorizations of gain-scaling matrices (cf.\ \cite[Section~7.3]{DaubSwel98}).  Theorem~\ref{thm:Semidirect} shows that the order-increasing property  implies~(\ref{trivial_gp}) and, by (\ref{unique_prod_rep}),  uniqueness of  product representations in (\ref{G=QK}).   Thus,   $\mathscr{S=D\ltimes C}$ implies uniqueness of product representations in $\mathscr{S=DC}$.  This is considerably weaker, however, than the  conclusion of~\cite[Theorem~1]{Bris:10:GLS-I}, which also follows from $\mathscr D$-invariance and the order-increasing property.   It is unclear whether~(\ref{trivial_gp}) follows from weaker assumptions than the nontrivial order-increasing property.

Let us reconcile uniqueness of product representations in  $\mathscr{S=DC}$ with nonuniqueness in~\cite[Theorem~1]{Bris:10:GLS-I}.  If $\mathbf{H}(z)$ has  multiple irreducible group lifting factorizations then \cite[Theorem~1]{Bris:10:GLS-I} says they are equivalent modulo rescaling; i.e.,
\begin{eqnarray*}
\mathbf{H}(z)\!\! & = &\!\! \mathbf{D}_K\,\mathbf{S}_{N-1}(z) \cdots\mathbf{S}_0(z)\, \mathbf{B}(z)  
			\,\equiv\,  \mathbf{D}\mathbf{E}\mathbf{B}\,\in\,\mathscr{S}\mathfrak{B}  \\
	&=&\!\!  \mathbf{D}_{K'}\,\gam{\alpha}\mathbf{S}_{N-1}(z)\cdots \gam{\alpha}\mathbf{S}_0(z)\,\mathbf{D}_{\alpha}\mathbf{B}(z) 
			\,\equiv\,  \mathbf{D}'\mathbf{E}'\mathbf{B}', 
\end{eqnarray*}
where $\alpha\equiv K/K'\neq 1$.  Since $\mathbf{B}'\equiv\mathbf{D}_{\alpha}\mathbf{B}\neq\mathbf{B}$ the $\mathscr{S}$-factors  are also different, $\mathbf{D'E'}\neq\mathbf{DE}$, so there is no contradiction with  uniqueness of product representations in $\mathscr{S=DC}$.

\subsubsection{Combining Theorems~\ref{thm:Cascade} and~\ref{thm:Semidirect}}
\label{sec:Semidirect:Scaled:Combine}
If $\mathfrak S$ is a $\mathscr D$-invariant, order-increasing group lifting structure then, by Lemma~\ref{lem:Cascade}, the hypotheses of both Theorem~\ref{thm:Cascade} and Theorem~\ref{thm:Semidirect} are satisfied.  Theorem~\ref{thm:Cascade} provides an isomorphism, call it $\rho$, that maps lifting matrices to tokens,
\[  \rho\colon\mathscr{C}\stackrel{\cong}{\longrightarrow}\mathscr{U \bigast L}.  \]
Theorem~\ref{thm:Semidirect} says that $\mathscr{S=D\ltimes C}$, and Lemma~\ref{lem:ext_semidirect} combines these representations into a single result. 

\begin{cor}\label{cor:S_structure}
If $\mathfrak S$ is a $\mathscr D$-invariant, order-increasing group lifting structure then its scaled lifting group has the structure
\begin{equation}\label{S_structure}
\mathscr{S} \cong\mathscr{D \ltimes_{\theta}(U\bigast L)}. 
\end{equation}
\end{cor}

\rem   The external semidirect product in Corollary~\ref{cor:S_structure} is based on the homomorphism $\theta$~(\ref{theta_automorphic}).  Let us make this abstractly defined homomorphism more concrete.  What is the action of the induced automorphism \thet{K} on reduced words, 
\[  \mathbf{E}=\mathbf{S}_N\cdots\mathbf{S}_0\in\mathscr{U\bigast L}? \]
It suffices to consider  individual tokens $\mathbf{S}_i$ and then combine these actions using the automorphism property,
\[  \thet{K}\mathbf{E} =  \thet{K}\mathbf{S}_N\cdots\thet{K}\mathbf{S}_0.  \]
Each token $\mathbf{S}_i$ is the image under  $\rho$ of some lifting matrix, $\mathbf{S}_i = \rho(\mathbf{S}_i(z))$, and  (\ref{def_theta})  says that the image of  $\mathbf{S}_i$ under  \thet{K} is
\[   \mathbf{S}'_i \equiv \thet{K}\mathbf{S}_i = \rho\left(\gam{K}\mathbf{S}_i(z)\right)=\rho\left(\mathbf{S}'_i(z)\right),  \]
where the corresponding lifting matrix  is given by the  inner automorphism
\[  \mathbf{S}'_i(z)\equiv \gam{K}\mathbf{S}_i(z) = \mathbf{D}_K\mathbf{S}_i(z)\mathbf{D}^{-1}_K.  \]
In other words, the action of \thet{K} on tokens corresponds to conjugation of lifting matrices by the scaling matrix $\mathbf{D}_K\in\mathscr{D}$.

\subsubsection{WS Filter Banks}\label{sec:Semidirect:Scaled:WS}
We can now give a group-theoretic characterization of the group of unimodular WS filter banks,  
\[  \mathscr{W=DC_W = S_W}.  \]
The  WS group lifting structure, 
$\mathfrak{S}_{\mathscr{W}}\equiv(\mathscr{D},\mathscr{U},\mathscr{L},\mathbf{I})$ \cite[Example~2]{Bris:10:GLS-I}, is  $\mathscr{D}$-invariant and  order-increasing \cite[Theorem~1]{Bris:10b:GLS-II} so  Corollary~\ref{cor:S_structure} implies the following.
\begin{cor}\label{cor:WS_classification}
Let $\mathfrak{S}_{\mathscr{W}}\equiv(\mathscr{D},\mathscr{U},\mathscr{L},\mathbf{I})$ be the group lifting structure for the unimodular WS  group, $\mathscr{W}$, defined in~\cite[Section~IV]{Bris:10:GLS-I}.  The group-theoretic structure of $\mathscr{W}$ is
\[  \mathscr{W} \cong\mathscr{D \ltimes_{\theta}(U\bigast L)}.  \]
\end{cor}

\subsubsection{HS Filter Banks}\label{sec:Semidirect:Scaled:HS}
We can also give a group-theoretic characterization of   $\mathfrak{H}$, the  class of all unimodular HS filter banks satisfying (\ref{HS_anal_mirror_DS}), even though it does not form a group.  The  group lifting factorization theory for $\mathfrak{H}$,
\begin{eqnarray}
\mathfrak{H} &=& \mathscr{DC_{\mathfrak{H}}} \mathfrak{B_H} = \mathscr{S}_{\mathfrak{H}} \mathfrak{B_H},\label{H_product_formula}\\
\mathfrak{B_H} &\equiv& \left\{\mathbf{B}\in\mathfrak{H}\colon\order(B_0)=\order(B_1)\right\}\label{B_HS},
\end{eqnarray}
only provides uniqueness modulo rescaling since $\mathfrak{B_H}$ is nontrivial \cite[Theorem~2]{Bris:10b:GLS-II}, but its group lifting structure is $\mathscr D$-invariant and order-increasing so Theorem~\ref{thm:Cascade} applies to $\mathscr{C}_{\mathfrak{H}}$ and Corollary~\ref{cor:S_structure} applies to $\mathscr{S}_{\mathfrak{H}}$.

The product representation (\ref{H_product_formula}) for $\mathfrak{H}$ has the form of a collection of right cosets of $\mathscr{S}_{\mathfrak{H}}$ by elements of $\mathfrak{B_H}$:
\begin{equation}\label{cosets_of_S}		
\mathfrak{H} = \mbox{$\bigcup$}\left\{\mathscr{S}_{\mathfrak{H}}\mathbf{B}\colon\mathbf{B}\in\mathfrak{B_H}\right\}.
\end{equation}
Distinct elements of $\mathfrak{B_H}$ do not generate distinct cosets of $\mathscr{S}_{\mathfrak{H}}$, however, because irreducible group lifting factorizations are only unique modulo rescaling.  To see this,  let $\mathbf{B}\in\mathfrak{B_H}$,  $\mathbf{H}=\mathbf{D}_{K}\mathbf{E}\mathbf{B}\in\mathscr{S}_{\mathfrak{H}}\mathbf{B}$, and $\alpha\neq 0$; then
\begin{equation}\label{alt_coset}
\mathbf{H}=\mathbf{D}_{\!K/\alpha}(\gam{\alpha}\mathbf{E})\mathbf{B}'\in\mathscr{S}_{\mathfrak{H}}\mathbf{B}',\quad
\mathbf{B}'\equiv\mathbf{D}_{\alpha}\mathbf{B}\in\mathfrak{B_H}.
\end{equation}
Since $\mathscr{S}_{\mathfrak{H}}\mathbf{B}\cap\mathscr{S}_{\mathfrak{H}}\mathbf{B}'\neq\emptyset$, a basic result in group theory~\cite[Proposition~III.19]{MacLaneBirkhoff67}, \cite[Corollary~I.4.3]{Hungerford74} says that these cosets are identical, $\mathscr{S}_{\mathfrak{H}}\mathbf{B}=\mathscr{S}_{\mathfrak{H}}\mathbf{B}'$.  

This coset duplication can be eliminated by  taking advantage of the fact that (\ref{B_HS}) is closed under scaling, i.e., that $\mathbf{D}_{K}\mathfrak{B_H}=\mathfrak{B_H}$.  Using (\ref{conjugation_operator}) and $\mathscr{D}$-invariance of $\mathscr{C}_{\mathfrak{H}}$, an arbitrary element of $\mathfrak{H}= \mathscr{S}_{\mathfrak{H}} \mathfrak{B_H}$ can be written
\begin{equation}\label{unique_coset_reps}
\mathbf{D}_K\mathbf{E}\mathbf{B}=(\gam{K}\mathbf{E})(\mathbf{D}_K\mathbf{B})
=\mathbf{E}' \mathbf{B}' \in\mathscr{C_{\mathfrak{H}}} \mathfrak{B_H} ,
\end{equation}
and  factorizations in $\mathscr{C_{\mathfrak{H}}} \mathfrak{B_H}$ are \emph{unique} because there are no gain-scaling matrices, so every $\mathbf{H}\in\mathfrak{H}$ is in a unique right coset $\mathscr{C_{\mathfrak{H}}} \mathbf{B}\subset\mathscr{C_{\mathfrak{H}}} \mathfrak{B_H}$.  

Alternatively, one can restrict $\mathfrak{B_H}$ to obtain unique group lifting factorizations.  If (\ref{B_HS}) is made more restrictive, e.g.,
\begin{equation}\label{B_HS2}
\mathfrak{B'_H}\equiv \left\{\mathbf{B}\in\mathfrak{B_H}\colon B_0(1)=1\right\},
\end{equation}
then $\mathbf{B}$ and $\mathbf{B}'\equiv\mathbf{D}_{\alpha}\mathbf{B}$ can both satisfy   (\ref{B_HS2}) only if $\alpha=1$.  Since any two irreducible group lifting factorizations of $\mathbf{H}$ in $\mathscr{S}_{\mathfrak{H}} \mathfrak{B_H}$ are equivalent modulo rescaling,  it follows that every $\mathbf{H}\in\mathfrak{H}$ is in a unique right coset  $\mathscr{S}_{\mathfrak{H}}\mathbf{B} \subset\mathscr{S}_{\mathfrak{H}} \mathfrak{B'_H}$.

Finally, \cite[Theorem~12]{BrisWohl06} implies that $\mathfrak{H}$ \emph{cannot} be expressed in terms of \emph{left} cosets  $\mathbf{B}\mathscr{S}_{\mathfrak{H}}$ or $\mathbf{B}\mathscr{C}_{\mathfrak{H}}$ for $\mathbf{B}\in\mathfrak{B_H}$.

\begin{cor}\label{cor:HS_classification}
Let $\mathfrak{S}_{\mathfrak{H}}\equiv(\mathscr{D},\mathscr{U},\mathscr{L},\mathfrak{B_H})$ be the group lifting structure for the unimodular HS  class, $\mathfrak{H}$, defined in~\cite[Section~IV]{Bris:10:GLS-I}.  Then the group-theoretic structure of $\mathscr{S}_{\mathfrak{H}}$ is
\begin{equation}\label{SH_group_structure}
\mathscr{S}_{\mathfrak{H}} \cong\mathscr{D \ltimes_{\theta}(U\bigast L)}.
\end{equation}
$\mathfrak{H}$ can be partitioned into disjoint right cosets  (but not left cosets) of either $\mathscr{C}_{\mathfrak{H}}$ or  $\mathscr{S}_{\mathfrak{H}}$, with $\mathfrak{B'_H}$  given by, e.g., (\ref{B_HS2}):
\begin{eqnarray}	
\mathfrak{H} & = & \mbox{$\bigcup$}\left\{\mathscr{C}_{\mathfrak{H}}\mathbf{B}\colon\mathbf{B}\in\mathfrak{B_H}\right\} \label{C_cosets}\\
		& = & \mbox{$\bigcup$}\left\{\mathscr{S}_{\mathfrak{H}}\mathbf{B}\colon\mathbf{B}\in\mathfrak{B'_H}\right\}.\label{S_cosets}
\end{eqnarray}
\end{cor}
%


\section{Comparison of Scaled Lifting Groups with Vector Spaces}
\label{sec:Comparison}
Finite-dimensional vector spaces are popular  parameter sets for numerical  design applications because every feasible solution has a unique representation as a linear combination of basis vectors.    Defining a vector space  framework for PR filter banks is problematic, however, since  filter banks naturally form nonabelian groups, not vector spaces.  In this section we  compare and contrast  the group-theoretic characterizations derived above, which  provide unique parametric factorizations   for scaled lifting groups of filter banks, with the more familiar unique factorization structures provided by vector spaces.  Throughout this section, $\mathscr{P}_0$ and $\mathscr{P}_1$ are finite-dimensional real vector spaces of lifting filters, such as filters of bounded orders satisfying (\ref{lowpass_HS}) and (\ref{highpass_HS}), respectively, with  upper and lower triangular lifting matrix groups  $\mathscr{U}\equiv\upsilon(\mathscr{P}_0)$ and $\mathscr{L}\equiv\lambda(\mathscr{P}_1)$.

\subsection{Unique Representations in Lifting Matrix Groups}\label{sec:Comparison:Unique}
We can write down a low-level,  homomorphic correspondence between vector space basis expansions of lifting filters and (abelian) matrix factorizations of individual lifting matrices.
Let $\left\{S_1,\ldots,S_{n_0}\right\}$ be a basis for the finite-dimensional vector space $\mathscr{P}_0$ of lifting filters for lowpass (upper triangular) lifting matrices. Every $S(z)\in \mathscr{P}_0 $ has a unique basis expansion 
\begin{equation}\label{P0_basis}
S(z) = \sum_{i=1}^{n_0} a_i S_i(z),\quad a_i\in\mathbb{R}.
\end{equation}
Similarly, every $T(z)\in \mathscr{P}_1 $ has a unique basis expansion in terms of a basis $\left\{T_1,\ldots,T_{n_1}\right\}$ for $\mathscr{P}_1$,
\begin{equation}\label{P1_basis}
T(z) = \sum_{i=1}^{n_1} a_i T_i(z),\quad a_i\in\mathbb{R}.
\end{equation}
These  expansions are transformed by the homomorphisms $\upsilon$, $\lambda$, and $\gamma$ into factorizations of  the corresponding lifting matrices that are unique  modulo permutations of the (commuting) matrix factors.  Let
\begin{eqnarray*}
 \kappa_i & \equiv & \sqrt{|a_i|}\quad\mbox{for $a_i\in\mathbb{R}$},\\
 \sigma_i & \equiv & \sgn(a_i) = \pm 1,\\
 \mathbf{S}_i(z) & \equiv & \upsilon(S_i(z)),\mbox{\ and}\\
 \mathbf{T}_i(z) & \equiv & \lambda(T_i(z)) .
 \end{eqnarray*}

With this notation, 
\[  \upsilon(a_i S_i(z)) = \upsilon(\sigma_i\kappa_i^2 S_i(z)) = \gamma_{\kappa_i}^{-1} \mathbf{S}_i^{\sigma_i}(z),  \]
where the inverse on $\gamma_{\kappa_i}^{-1}=\gamma_{\kappa_i^{-1}}$ is required by (\ref{gain_matrix}) and (\ref{conjugation_operator}) in the upper triangular case.  The isomorphism $\upsilon$ thus transforms the basis expansion (\ref{P0_basis}) for a lifting filter in $\mathscr{P}_0$ into a unimodular matrix factorization of the corresponding lifting matrix in $\mathscr{U}\equiv\upsilon(\mathscr{P}_0)$,
\begin{eqnarray}
\upsilon(S(z)) &=& \upsilon(a_1 S_1(z))\cdots\upsilon(a_{n_0} S_{n_0}(z))\nonumber\\
	&=& \gamma_{\kappa_1}^{-1}\mathbf{S}_1^{\sigma_1}(z)\cdots\gamma_{\kappa_{n_0}}^{-1}\mathbf{S}_{n_0}^{\sigma_{n_0}}(z).\label{U_basis}
\end{eqnarray}
The isomorphism $\lambda$ similarly transforms (\ref{P1_basis}) into a lower triangular matrix factorization in  $\mathscr{L}\equiv\lambda(\mathscr{P}_1)$,
\begin{eqnarray}
\lambda(T(z)) &=& \lambda(a_1 T_1(z))\cdots \lambda(a_{n_1} T_{n_1}(z))\nonumber\\
&=& \gamma_{\kappa_1} \mathbf{T}_1^{\sigma_1}(z)\cdots\gamma_{\kappa_{n_1}} \mathbf{T}_{n_1}^{\sigma_{n_1}}(z).\label{L_basis}
\end{eqnarray}

Formulas (\ref{U_basis}) and (\ref{L_basis}) are ``basis expansions'' for $\mathscr{U}$ and $\mathscr{L}$.  Uniqueness of the parameters $a_i$ in  (\ref{P0_basis}) and (\ref{P1_basis}) implies uniqueness of the parameters $\kappa_i$ and $\sigma_i$ in (\ref{U_basis}) and (\ref{L_basis}).
This furnishes unique parametric  factorizations for lower and upper triangular lifting matrices with lifting filters drawn from finite-dimensional vector spaces of polynomials. 

Note that uniqueness of the lifting matrix factorizations~(\ref{U_basis}) and~(\ref{L_basis}) has nothing to do with unique factorization properties of group lifting structures; it is a simple consequence of uniqueness of basis expansions in the underlying vector spaces of lifting filters. If, however, a group lifting structure incorporating these two lifting matrix groups is $\mathscr{D}$-invariant and order-increasing then every filter bank  in $\mathscr{S}$ has a unique \emph{irreducible} group lifting factorization.   The group-theoretic structure $\mathscr{S} \cong\mathscr{D \ltimes_{\theta}(U\bigast L)}$ given in  Corollary~\ref{cor:S_structure} therefore provides a vector space-like unique factorization framework for members of the scaled lifting group in terms of ``basis elements''~(\ref{U_basis}) and~(\ref{L_basis}), with  ``scalar multiplication'' given by unimodular scaling matrices in a gain-scaling group, $\mathscr{D}$.

\subsection{Automorphic Scaling Operations}\label{sec:Comparison:Scaling}
Next, we explore the parallels between scaled lifting groups and vector spaces more closely. Assume we have been given a  $\mathscr{D}$-invariant, order-increasing group lifting structure, where the gain-scaling group $\mathscr{D}\equiv\mathbf{D}(\mathscr{R})\cong\mathscr{R}$ is the isomorphic image of a multiplicative group $\mathscr{R}$ of  real  numbers (\ref{D_iso}).   Recall the axioms for vector spaces~\cite{Strang:88:Linear-Algebra-Applications,MacLaneBirkhoff67,Hungerford74,Jacobson74}.
\begin{defn}[Vector space]\label{defn:VectorSpace}
A \emph{vector space} is an abelian group, $(V,+)$, together with a field $\mathbb F$ and a scalar multiplication operation  that satisfies the following axioms for all $a,\,b\in\mathbb F$ and $\bsy{u},\,\bsy{v}\in V$.
\begin{eqnarray}
(a+b)\bsy{v} &=& a\bsy{v}+b\bsy{v}\label{dist_over_scalar_add}\\
a(\bsy{u}+\bsy{v}) &=& a\bsy{u}+a\bsy{v}\label{dist_over_vector_add}\\
a(b\bsy{v}) &=& (ab)\bsy{v}\label{mult_assoc} \\
1_{\ssst\mathbb F}\bsy{v} &=& \bsy{v}\label{mult_id}
\end{eqnarray}
\end{defn}

We shall now show that axioms~(\ref{dist_over_vector_add}), (\ref{mult_assoc}), and (\ref{mult_id}) all have multiplicative (homomorphic) analogues in $\mathscr S$ using the automorphic scaling action $\gamma\colon\mathscr{D}\rightarrow\Aut(\mathscr{S})$ but that axiom~(\ref{dist_over_scalar_add}) does \emph{not}, at least under one fairly reasonable interpretation of what a homomorphic analogue of axiom~(\ref{dist_over_scalar_add}) would be, no matter how we might try to redefine automorphic scaling.

Axiom~(\ref{dist_over_vector_add}) says that scalar multiplication distributes over the group operation in $V$.  The homomorphic analogue in  $\mathscr{S}$, where the group operation is matrix multiplication, is 
\begin{equation}\label{scaling_homo}
\gam{K}(\mathbf{EF}) = (\gam{K}\mathbf{E})(\gam{K}\mathbf{F}),
\end{equation}
which is just the automorphism property of $\gam{K}\colon\mathscr{S}\rightarrow\mathscr{S}$.

Axiom~(\ref{mult_assoc}) is an associative law for  scaling.  This translates into  the homomorphism property of $\gamma\colon\mathscr{D}\rightarrow\Aut(\mathscr{S})$,
\begin{equation}\label{scaling_composition}
\gam{K'}(\gam{K}\mathbf{E}) = \gam{K'K}\mathbf{E},
\end{equation}
which says that $\gamma$ is a group action of $\mathscr{D}$ on $\mathscr{S}$.
 
Axiom~(\ref{mult_id}) says that the multiplicative unit element acts as the identity operator.  Its analogue in $\mathscr S$ is the fact that the homomorphism $\gamma$ maps $\mathbf{D}_1=\mathbf{I}$ to the identity automorphism:
\begin{equation}\label{scaling_identity}
\gamma_{\ssst 1}\mathbf{E} = \mathbf{E}.
\end{equation}

What about Axiom~(\ref{dist_over_scalar_add}), which says that scalar multiplication distributes over scalar addition?  Finding a homomorphic analogue is complicated by the fact that the multiplicative group  $\mathscr D$ doesn't necessarily have an additive structure, so suppose $\mathscr{R}$ is closed under addition, e.g., \mbox{$\mathscr{R}=(0,\infty)$}, the positive real numbers.  As in (\ref{scaling_homo}), we assume that the group operation (vector addition) on the  right-hand side of (\ref{dist_over_scalar_add}) is mapped  to the group operation in $\mathscr{S}$ (matrix multiplication).  If we regard \mbox{$K\mapsto\gam{K}$} as a mapping of $\mathscr{R}$ into $\Aut(\mathscr{S})$,  can automorphic scaling in $\mathscr{S}$ distribute over  addition in $\mathscr{R}$, i.e., can \mbox{$K\mapsto\gam{K}\mathbf{E}$} be an additive homomorphism:
\begin{equation}\label{scaling_add}
\gam{K+K'}\mathbf{E} = (\gam{K}\mathbf{E})(\gam{K'}\mathbf{E})\,?
\end{equation}

While (\ref{scaling_add}) is not satisfied by the scaling action $\gamma$ defined by (\ref{gain_matrix}) and (\ref{conjugation_operator}), is there any way to \emph{redefine} gain-scaling (e.g., using exponential functions) to make $\gamma$ satisfy (\ref{scaling_add})?  If so, \mbox{$K+K'=K'+K$} implies that $\gam{K}\mathbf{E}$ and $\gam{K'}\mathbf{E}$ would  have to commute for all $\mathbf{E}\in\mathscr{S}$ and $K',K\in\mathscr{R}$.  In fact, something even stronger would have to be true.

\begin{prop}\label{prop:AdditivityImpliesCommutativity}
Let the multiplicative group $\mathscr{R}$ be closed under addition.  If there exists a multiplicative homomorphism  \mbox{$\gamma\colon\mathscr{R}\rightarrow\Aut(\mathscr{S})$} that also satisfies (\ref{scaling_add})  for all $\mathbf{E}\in\mathscr{S}$ and $K',K\in\mathscr{R}$ then $\mathscr{S}$ is  abelian.
\rem
For any nonabelian group $\mathscr{S}$, a multiplicative homomorphism  \mbox{$\gamma\colon\mathscr{R}\rightarrow\Aut(\mathscr{S})$} automatically satisfies (\ref{scaling_homo}), (\ref{scaling_composition}), and (\ref{scaling_identity}).
In the language of abstract algebra, however, Proposition~\ref{prop:AdditivityImpliesCommutativity} says there is no such thing as a ``nonabelian module'' that also satisfies (\ref{scaling_add}).
\pf
Let $\mathbf{E,F}\in\mathscr{S}$. Since $\mathscr{R}$ is also closed under addition, $2=1+1\in\mathscr{R}$.  The homomorphism $\gamma$ maps 1 to the identity automorphism $\gamma_{\ssst 1}$ so  (\ref{scaling_add}) implies
\begin{equation}\label{scaling_add_consequence1}
\gamma_{\ssst 2}\mathbf{E}  =  (\gamma_{\ssst 1}\mathbf{E})^2  =  \mathbf{E}^2
\end{equation}
and
\begin{equation} \label{scaling_add_consequence2}
\gamma_{\ssst 2}(\mathbf{EF}) = (\mathbf{EF})^2 .
\end{equation}
On the other hand,  $\gamma_{\ssst 2}\in\Aut(\mathscr{S})$ and (\ref{scaling_add_consequence1}) imply that
\begin{equation} \label{scaling_add_consequence3}
\gamma_{\ssst 2}(\mathbf{EF}) = (\gamma_{\ssst 2}\mathbf{E})(\gamma_{\ssst 2}\mathbf{F}) = 
	\mathbf{E}^2\mathbf{F}^2 .
\end{equation}
Equate (\ref{scaling_add_consequence2}) and (\ref{scaling_add_consequence3}) and cancel common factors:
\begin{eqnarray*}
(\mathbf{EF})^2 &=& \mathbf{E}^2\mathbf{F}^2 \\
\mathbf{FE} &=& \mathbf{EF}.
\end{eqnarray*}
\hfill\qed
\end{prop}

Since scaled lifting groups are  nonabelian, (\ref{scaling_add}) fails for \emph{any} automorphic scaling operation.  We have thus shown that \mbox{$\mathscr{S=D\ltimes C}$}  has a scaling structure that  is \emph{partially} homomorphic to scalar multiplication in vector spaces.  

This phenomenon has precedent, and other continuous groups with scaling automorphisms have been studied in the literature.  For instance, \emph{homogeneous groups} \cite{FollandStein:82:Hardy-Spaces,Stein:93:Harmonic-Analysis} are  nilpotent Lie groups equipped with dilations (families of automorphisms $\delta_r,\;r>0$, that act as dilations on local coordinates for the group).   The class of homogeneous groups, which includes the Heisenberg group, has attracted  attention because of its close connections to harmonic analysis, mathematical physics, and partial differential equations.  Unfortunately, scaled lifting groups of the form $\mathscr{D \ltimes_{\theta}(U\bigast L)}$ are not nilpotent, so we  leave the connection between  scaled lifting groups  and other  continuous groups-with-dilations as an open question.


\section{Conclusions}\label{sec:Conclusions}
The growing importance of multirate filter banks in  digital communication standards, combined with the fact that  filter banks do not form vector spaces, has convinced the author that a better understanding of the field can be gained by  employing some well-established tools from algebraic group theory.  The   structure theory derived here for groups of linear phase filter banks provides a mathematical framework containing homomorphic analogues of many familiar linear algebraic properties.  It is hoped that the detailed parameterizations of linear phase filter banks described in this paper will prove useful for filter bank designs based on parametric numerical optimization since the above classification is both complete and injective; i.e., a given filter bank is not encountered at multiple points in parameter space by  optimization algorithms. 

The lifting cascade group and scaled lifting group generated by a $\mathscr{D}$-invariant, order-increasing group lifting structure, $\mathfrak{S}=(\mathscr{D,U,L,}\mathfrak{B})$, have been determined up to isomorphism in terms of the building blocks $\mathscr{U,\,L,}$ and $\mathscr{D}$.  The unique factorization theorem for $\mathscr{D}$-invariant, order-increasing group lifting structures in \cite{Bris:10:GLS-I} implies that  the lifting cascade group $\mathscr{C\equiv\langle U\cup L\rangle}$ is isomorphic to the free product, $\mathscr{U \bigast L}$, of the abelian lifting matrix groups $\mathscr{U}$ and $\mathscr{L}$ (Theorem~\ref{thm:Cascade}).  It  is shown that $\mathscr{C}$ is a free group if and only if  $\mathscr{U}$ and $\mathscr{L}$ are infinite cyclic groups and  $\mathscr{C\cong U \bigast L}$ (Theorem~\ref{thm:FreeCascadeGroup}).  

When $\mathfrak{S}$ is $\mathscr{D}$-invariant and order-increasing it has also been shown that the  scaled lifting group $\mathscr{S}\equiv\langle\mathscr{C\cup D}\rangle$ is given by the internal semidirect product of $\mathscr{C}$ by $\mathscr{D}$ (Theorem~\ref{thm:Semidirect}).  This result is proven in a relatively simple way directly from the $\mathscr{D}$-invariance and order-increasing hypotheses without explicitly invoking uniqueness of irreducible group lifting factorizations.  Combining Theorems~\ref{thm:Cascade} and~\ref{thm:Semidirect} characterizes  $\mathscr{D}$-invariant, order-increasing scaled lifting groups up to isomorphism in terms of $\mathscr{U,\,L,}$ and $\mathscr{D}$ (Corollary~\ref{cor:S_structure}):
\[ \mathscr{S} \cong\mathscr{D \ltimes_{\theta}(U\bigast L)}. \]

This result applies to the group $\mathscr{W=S_W}$ of unimodular whole-sample symmetric filter banks  specified in JPEG~2000 Part~2 Annex~G (Corollary~\ref{cor:WS_classification}).  It also applies to the scaled lifting group $\mathscr{S}_{\mathfrak{H}}$ for the unimodular half-sample symmetric class, $\mathfrak{H}$.  While $\mathfrak{H}$ does \emph{not} form a group it can be partitioned into cosets of either $\mathscr{S}_{\mathfrak{H}}$ or $\mathscr{C}_{\mathfrak{H}}$  (Corollary~\ref{cor:HS_classification}).  Homomorphic comparisons are made between basis expansions in vector spaces and the unique factorization structure of scaled lifting groups for $\mathscr{D}$-invariant, order-increasing group lifting structures.  It is shown that such scaled lifting groups can be regarded as noncommutative multiplicative analogues of vector spaces.


\section*{Acknowledgments}
The author thanks Associate Editor Olgica Milenkovic and two anonymous reviewers for their constructive input.  He also thanks the producers of the Ipe drawing editor (\texttt{http://ipe7.sourceforge.net}) and  the \TeX Live/Mac\TeX\ distribution (\texttt{http://www.tug.org/mactex}).



\begin{IEEEbiographynophoto}
{Christopher M.\ Brislawn} (M'91--SM'05) received the B.S.\ degree in 1982 from Harvey Mudd College, Claremont, CA, and the Ph.D.\ degree in 1988 from the University of Colorado---Boulder, both in mathematics.

He was a Visiting Assistant Professor  at the University of Southern California from 1989 to 1990 and joined Los Alamos National Laboratory (LANL), Los Alamos, NM,  as a postdoc in 1990.  Currently he is a  Scientist in Group CCS-3, Information Sciences, in the Division of Computer, Computational and Statistical Sciences at LANL.  From 1991 to 1993 he coauthored the Wavelet/Scalar Quantization Specification for compression of digitized fingerprint images with the U.S.\ Federal Bureau of Investigation.  From 1999 to 2003 he served as LANL's Principal Member of Working Group L3.2 on the International Committee for Information Technology Standards and led a LANL team that worked on the ISO/IEC JPEG~2000 standard (ISO 15444-x).  He served as the first editor of JPEG~2000 Part~10 (Extensions for Three-Dimensional Data).  In 2007--2008 he represented LANL on  the Motion Imagery Standards Board for the National Geospatial Intelligence Agency.  He has mentored numerous graduate students and postdocs at LANL and has co-supervised a Ph.D.\ dissertation for the University of Texas---Austin.  His current research interests include multirate filter banks and wavelet transforms,  communications coding, digital signal and image processing,  network signal processing,  joint time-frequency analysis, and numerical linear algebra.  

Dr.\ Brislawn is also a member of the American Mathematical Society.
\end{IEEEbiographynophoto}

\end{document}